\documentclass[%
superscriptaddress,
preprintnumbers,
showkeys,
nofootinbib,
 amsmath,amssymb,
 aps,
prd,
floatfix,
]{revtex4} % {article} % {revtex4-1} 
\pdfoutput=1
\usepackage{float}
\usepackage{nicefrac}
\usepackage{slashed}
\usepackage{mathtools}
\usepackage{amsfonts} % AMS
\usepackage{amssymb} % AMS
\usepackage{amsmath} % AMS
\usepackage{dsfont}
\usepackage{esint}
\usepackage{graphicx} % Include figure files
\usepackage{subfigure} % Include figure files
\usepackage{array} % array
\usepackage{dcolumn} % Align table columns on decimal point
\usepackage{bm} % bold math
\usepackage{latexsym} % latex symbols
\usepackage{longtable} % long tables
\usepackage{hyperref} % hypertext links 
\usepackage{verbatim}
\usepackage{epsfig}
\usepackage[usenames,dvipsnames,svgnames,table]{xcolor}
\begin{document}
\title{From deep inelastic scattering to heavy-flavor semi-leptonic decays:\\Total rates into multi-hadron final states from lattice QCD}

%%%%%%%%%%
\author{Maxwell T.\ Hansen}
\email[e-mail: ]{maxhanse@uni-mainz.de}
\affiliation{Helmholtz~Institut~Mainz, D-55099 Mainz, Germany.\\[3pt]}
%%%%%%%%%%

%%%%%%%%%%
\author{Harvey B.\ Meyer}
\email[e-mail: ]{meyerh@kph.uni-mainz.de}
\affiliation{Helmholtz~Institut~Mainz, D-55099 Mainz, Germany.\\[3pt]}
\affiliation{PRISMA Cluster of Excellence and Institut f\"ur Kernphysik,\\ 
Johannes Gutenberg-Universit\"at Mainz, D-55099 Mainz, Germany.\\[3pt]}
%%%%%%%%%%

%%%%%%%%%%
\author{Daniel Robaina}
\email[e-mail: ]{robaina@theorie.ikp.physik.tu-darmstadt.de}
\affiliation{Institut f\"ur Kernphysik, Technische Universit\"at Darmstadt,
Schlossgartenstrasse 2, D-64289 Darmstadt, Germany.}
%%%%%%%%%%

%%%%%%%%%%
\date{\today}
%%%%%%%%%%

%%%%%%%%%%
\begin{abstract}
We present a new technique for extracting decay and transition rates into final states with any number of hadrons. The approach is only sensitive to total rates, in which all out-states with a given set of QCD quantum numbers are included. For processes involving photons or leptons, differential rates with respect to the non-hadronic kinematics may also be extracted. Our method involves constructing a finite-volume Euclidean four-point function, whose corresponding spectral function measures the decay and transition rates in the infinite-volume limit. This requires solving the inverse problem of extracting the spectral function from the correlator and also necessitates a smoothing procedure so that a well-defined infinite-volume limit exists. Both of these steps are accomplished by the Backus-Gilbert method and, as we show with a numerical example, reasonable precision can be expected in cases with multiple open decay channels. Potential applications include nucleon structure functions and the
onset of the deep inelastic scattering regime, as well as
semi-leptonic $D$ and $B$ decay rates. 
 \end{abstract}
%%%%%%%%%%

\preprint{MITP/17-031}

\nopagebreak

\maketitle

\section{Introduction\label{sec:intro}}

Reliably calculating the low-energy phenomenology of the strong interaction is very challenging. One major source of complication is that the underlying gauge theory, quantum chromodynamics (QCD), is most simply expressed in terms of color-charged quarks and gluons, whereas the low-energy degrees of freedom of the theory are confined color-singlet states, called hadrons. In the past few decades, great progress has been made in extracting the low-energy hadron spectrum and its structure directly from QCD, by numerically estimating the quantum path integral on a discretized, finite-volume Euclidean space-time. 

This approach, called lattice QCD, enables one to calculate the discrete spectrum of QCD on a three-dimensional torus, i.e.~a cubic spatial volume with periodicity $L$, as well as matrix elements of quark-field operators between finite-volume hamiltonian eigenstates. Taking discretization and finite-time effects to be negligible, it is important to understand how the energies and matrix elements on the torus can be related to the infinite-volume observables of the theory. 

For example, the energies of single-particle states with vanishing spatial momentum are known to satisfy $E_{\mathcal Q}(L) = M_{\mathcal Q} + \mathcal O(e^{- M_\pi L})$, where $M_\pi$ is the physical mass of the lightest degree of freedom, the pion in QCD, and $M_{\mathcal Q}$ the mass of the particle with quantum numbers ${\mathcal Q}$ \cite{LuscherStable1986}. For such states, one aims to calculate $E_{\mathcal Q}(L)$ at multiple volumes and extrapolate to the infinite-volume limit. Similarly, local matrix elements with single-hadron states are exponentially close to their infinite-volume counterparts, so that one can hope to estimate the experimental observable by taking $M_\pi L$ sufficiently large.

By contrast, for multi-particle states, a naive infinite-volume limit is not useful. For a given box size, the energy of the $n^{\rm th}$ multi-particle state contains information about the interactions of the particles, but as $L \to \infty$ with $n$ fixed, this state flows to the threshold value, $2M_{\mathcal Q}$. Instead, the now-standard approach is to use finite-volume as a tool, rather than an unwanted artifact, by applying analytic, field-theoretic relations between finite- and infinite-volume quantities. 

This approach was pioneered by L\"uscher, who derived a quantization condition relating the finite-volume spectrum to the two-to-two scattering amplitude in the case of identical scalar particles \cite{Luscher1986,Luscher1990}. The formalism includes all partial waves but assumes that the two-particle states have zero total momentum in the finite-volume frame and also neglects exponentially suppressed terms of the form $e^{- M_\pi L}$. More recently, this idea has been generalized to describe systems with nonzero momentum, as well as any number of strongly-coupled two-particle channels, including non-identical and non-degenerate particles as well as intrinsic spin~\cite{Rummukainen1995,KSS2005,Christ2005,Lage2009,Bernard2010,Doring2011,Fu2012,Gockeler2012,HSmultiLL,BricenoTwoPart2012,BricenoSpin}. Extensions to three-particle scattering are also underway~\cite{LtoK,KtoM,2to3}.

\bigskip

A similar approach can be used to relate matrix elements of finite-volume multi-particle states to their infinite-volume counterparts, in order to extract electroweak decay and transition amplitudes. Here the paradigm example is the weak decay $K \to \pi \pi$. The decay amplitude is given at leading order in the weak interaction by the QCD matrix element $\langle \pi \pi, \mathrm{out} \vert \mathcal H(0) \vert K \rangle$, where $\mathcal H(0)$ is the weak hamiltonian density expressed in terms of four-quark operators. In lattice QCD, however, it is only possible to calculate $\langle n, L \vert \mathcal H(0) \vert K, L \rangle$ where $\vert n, L \rangle$ is a finite-volume state with the quantum numbers of two pions.

In Ref.~\cite{Lellouch2000}, Lellouch and L\"uscher derived a conversion factor relating these finite- and infinite-volume matrix elements. The factor depends on the box size via a known geometric function and also depends on the derivative of the elastic pion scattering phase shift, $\partial \delta_{\pi \pi}(k)/\partial k$, evaluated at the kaon mass. This relation is being used by the RBC/UKQCD collaboration to perform a full-error-budget calculation of $K \to \pi \pi$ decays, aiming towards a first-principles understanding of the $\Delta I=1/2$ rule and a prediction of $\epsilon'/\epsilon$~\cite{KpipiThreeHalves2015,KpipiCPV2015}.

The Lellouch-L\"uscher relation has since been generalized to states with non-zero total momentum in the finite volume and to coupled two-particle channels with non-identical and non-degenerate particles, as well as particles with intrinsic spin \cite{KSS2005,Christ2005,MeyerTimelike2011,HSmultiLL,BricenoTwoPart2012,BHWOneToTwo,BHOneToTwoSpin}. These extensions also accommodate currents that carry angular-momentum, momentum and energy, allowing one to extract time-like form factors as well as semi-leptonic decay amplitudes. In Ref.~\cite{BH2to2} the relation was also extended to matrix elements of the form $\langle \pi \pi, \mathrm{out} \vert \mathcal J(0) \vert \pi \pi, \mathrm{in}\rangle$, providing a rigorous path towards resonance form factors and transition amplitudes. Going beyond single-current insertions, Ref.~\cite{Christ:2015pwa} used techniques based in the Lellouch-L\"uscher approach to analyze long-distance contributions to $K_L$-$K_S$ mixing from a finite-volume Euclidean four-point function.

Aside from the two-to-two transitions and the neutral meson mixing formalism, all relations of the Lellouch-L\"uscher type have the general form
\begin{equation}
\label{eq:genLL}
\langle E_k(L), \textbf p \, \vert \mathcal J (0) \vert \Psi, \textbf P \rangle_L = \sum_{\alpha} \mathcal C_\alpha \  \langle E_k(L),  \textbf p , \alpha,  \mathrm{out} \vert \mathcal J (0) \vert \Psi, \textbf P \rangle \,,
\end{equation}
where $\vert \Psi, \textbf P \rangle$ is a QCD-stable, single-particle state with the indicated three-momentum. Here $\alpha$ is summed over all two-particle channels with the relevant quantum numbers. For example, in the case of unflavored states, the sum includes $\pi \pi$ and $K \overline K$. In words, the finite-volume matrix element on the left-hand side of Eq.~(\ref{eq:genLL}) is equal to a linear combination of all possible infinite-volume matrix elements whose out-states have the appropriate quantum numbers. The coefficients, $\mathcal C_\alpha$, depend on derivatives of all parameters in this sector of the QCD scattering matrix as well as on the box size.

Using this result to extract decay and transition amplitudes for heavy mesons is extremely challenging. For a system with $N$ open decay channels, one must first use extensions of the L\"uscher formalism to extract all QCD scattering parameters, at multiple energies near the target decay energy, in order to estimate the derivatives. Given these, it is possible to calculate the coefficients, $\mathcal C_\alpha$. In a second step, one must then calculate finite-volume matrix elements at different box sizes but with the same energies, in order to generate multiple equations of the form~(\ref{eq:genLL}), with only the infinite-volume matrix elements unknown. Given $N$ independent results one can invert the relations to determine the various transition amplitudes.\footnote{In certain cases an ambiguity in the overall phase may require an additional constraint beyond the $N$ that are always needed, see also the discussion in Ref.~\cite{HSmultiLL}.}

It is also important to note that this approach is currently only available for energies below the production of the lightest state with more than two hadrons. For example, in the case of unflavored final states, one is limited by either the three- or four-pion threshold. The situation is particularly frustrating for the decay of charmed or bottom mesons as these can produce pions copiously, and only by disentangling all open channels can one make a statement, about, say, $D \to \pi \pi, K \overline K$. In addition, we note that the number of open channels, $N$, counts not only the species in the asymptotic states but also the angular momentum. For example, in the case of $N \gamma \to N \pi$ transitions, the finite-volume matrix element will be contaminated by multiple angular-momentum states, a problem that worsens as the
center-of-mass energy is increased.

\bigskip

Given these complications, it is prudent to investigate whether an alternative approach can be used to extract decay and transition rates from lattice QCD. In this work, we consider cases in which total transition rates into all out-states with given QCD quantum numbers are of interest. This is less information than the individual transition amplitudes of Eq.~(\ref{eq:genLL}), and it is reasonable to ask whether it can be accessed in a more direct manner.

Our method is inspired by Fermi's Golden Rule. Imagine a quantum system defined initially on a three-dimensional torus, described by a Hamiltonian $H=H_0+V$,
with an unperturbed part $H_0$ under which a particle $\Psi$ is stable (with mass $M$),
and a small perturbation $V$ that allows it to decay.
According to the standard derivation of Fermi's Golden Rule, the decay rate
of $\Psi$ in its rest frame in infinite volume can be calculated using the 
double limit\footnote{To be precise, the $k$th term in Eq.\ (\ref{eq:FGR}) is 
equal to the probability that in a measurement done at time $t$ the system
is observed in the unperturbed state $k$, divided by $t$.}
\begin{equation}
\label{eq:FGR}
\Gamma = 2\pi \lim_{t\to\infty}\lim_{L\to\infty} \sum_{k} |V_k(L)|^2\, \delta_{\nicefrac{1}{t}}(E_k(L)-M),
\end{equation}
where $V_k(L) = \langle k, L |V|\Psi \rangle $ 
is a finite-volume transition matrix element with unit-norm states and where
\begin{equation}
\delta_{\nicefrac{1}{t}}(\omega) = \frac{2}{\pi} \frac{\sin^2(\omega t/2)}{\omega^2 t} \,,
\end{equation}
can be interpreted as a regularized delta-function.
The sum extends over states whose energy $E_k(L)$ lies within a fixed interval centered at $M$.
A key observation that we exploit in this paper is that, if one is only interested in $\Gamma$,
other forms of regularized delta-functions can be inserted into 
Eq.\ (\ref{eq:FGR}) without affecting the result, as long as the regularization is removed at the end.
Note that the order of limits in Eq.\ (\ref{eq:FGR}) is important.

We can rewrite the particle width as 
\begin{eqnarray}\label{eq:FGR2}
\Gamma &=&  \frac{1}{2M}\lim_{\Delta\to0}\lim_{L\to\infty} \int_0^\infty d\omega\, 
\Big[  4\pi M\sum_{k=0}^\infty |V_k(L)|^2\, \delta(\omega-E_k(L)) \Big]\;
\widehat\delta_\Delta(M,\omega)\, ,
\end{eqnarray}
where the \emph{resolution function} 
$\widehat\delta_\Delta(M,\omega)$ is a regularized delta-function centered at energy $M$ with a 
characteristic width given by $\Delta$ and falling off sufficiently rapidly at large $\omega$. The expression in square brackets is the finite-volume spectral function, denoted $\rho(\omega,L)$.
Related spectral functions are a central concept in finite-temperature lattice QCD studies 
(see e.g.\ \cite{Meyer:2011gj,Aarts:2014nba,Brandt:2015aqk,Ghiglieri:2016tvj,Burnier:2013nla}\footnote{In that context, 
the projector onto the initial state $|\Psi \rangle \langle \Psi|$ is replaced by the canonical
thermal average $\frac{1}{Z}\sum_n e^{-\beta E_n}|n  \rangle  \langle n |$.}).  Bringing
these ideas together, in this work we present a formalism for
extracting total decay and transition rates by extracting smeared
spectral functions from appropriately constructed finite-volume,
Euclidean correlation functions. The formalism is valid for
final-states with any number of outgoing hadrons and does not require
disentangling the exclusive transition amplitudes.

Our method requires solving an inverse problem of the Laplace type
\begin{equation}
G(\tau,L) =  \int_0^\infty  \frac{d \omega}{2 \pi} \; e^{-\omega \tau} \,\rho(\omega,L) \,,
\end{equation}
where $G(\tau,L)$ is a Euclidean correlator computed on the lattice.
We envision using the Backus-Gilbert method~\cite{Backus1,Backus2,Press:2007zz,Brandt:2015sxa,Brandt:2015aqk} to achieve this aim. This gives an estimate for the finite-volume spectral function, smeared by a resolution function that regulates the delta-functions corresponding to the discrete finite-volume energies. The smeared spectral function has precisely the form appearing in the double limit of Eq.~(\ref{eq:FGR2}). By varying the width of the resolution function and the box size of the calculation, one can estimate the double limit required to extract total decay or transition rates.

We emphasize that, in processes where the hadronic final states are produced
with varying energy, such as in semileptonic $D$ or
$B$ decays, the smeared spectral function obtained from the lattice
can be compared model-independently to
the experimental decay rate, at the cost of smearing the experimental
information in the exact same way. This comparison may be performed in a model independent way with the Backus-Gilbert method, because the
smearing kernel (i.e.\ the resolution function) is known exactly.  In
particular, no \emph{ad hoc} functional ansatz is required for the spectral
function.

We note that the present application has two distinct advantages over
the finite-temperature studies that use the same method. First, the
larger range of Euclidean times in zero-temperature
calculations should lead to better constraints on the corresponding
spectral function. Second, the Backus-Gilbert method is known to work
best for slowly varying spectral functions and to struggle in
determining narrow features such as resonance peaks. In the present
case, we are interested in the spectral function near the energy of the
mother particle, but for differing QCD quantum numbers, i.e.~those of
the final states after the current-mediated transition. Thus, one can
hope that in many instances the spectral function will not exhibit a rapid
variation, in which case little information is lost due to our lack 
of energy-resolution.

We close the introduction by highlighting some additional %two other 
techniques for studying multi-hadron states, beyond the methods related to that of Lellouch and L\"uscher. First, Ref.~\cite{Meissner:2010rq} describes an idea for extracting resonance parameters directly from Euclidean two-point functions. The approach is applicable to systems with a well-isolated, low-lying narrow resonance and makes use of a model-independent parameterization that allows one to determine resonance parameters by performing a fit to numerical lattice data. Like our approach, the technique of Ref.~\cite{Meissner:2010rq} requires estimating the $L \to \infty$ limit. 

Second, Ref.~\cite{Agadjanov:2016mao} gives a method for extracting the optical potential from a numerical lattice calculation. This allows one to access a particular scattering channel, for example $K \overline K$, above the threshold of another channel, for example $\pi \pi$. The finite-volume optical potential is known to contain an infinite tower of poles, and one must fit to a function of this form, apply an $i \epsilon$ prescription, and finally estimate the limit $L \to \infty$ followed by $\epsilon \to 0$. The $i \epsilon$ provides an effective smearing of the optical potential that makes the infinite-volume limit well defined, analogous to the smearing that we achieve via the Backus-Gilbert method. However, the method of Ref.~\cite{Agadjanov:2016mao} differs, for example, in that it relies on extracting finite-volume energy levels and requires twisted boundary conditions to sample the finite-volume optical potential.

Third, Refs.~\cite{Liu:1993cv,Liu:2016djw,Liu:2017lpe} discuss prospects for studying the hadronic tensor by solving the inverse problem on a Euclidean four-point function in the same way as we describe in this work. In contrast to the present study these references do not advocate the Backus-Gilbert technique and do not discuss finite-volume effects, the need for smoothing and the ordered double limit that plays a central role in the present analysis. 

Fourth, and finally, Ref.~\cite{Hashimoto:2017wqo} proposes to use, and performs an exploratory lattice study of Euclidean four-point functions to study semi-leptonic B decays. Again, the main contrast to our work is that the role of the finite volume is not discussed. Ref.~\cite{Hashimoto:2017wqo} also advocates avoiding the inverse problem by instead integrating the experimental data against a multi-pole function to extract tailored moments that can be more directly compared to lattice data. 

%Finally we would like to highlight Ref.~\cite{Lin:2001ek}, which provides an alternative derivation of the Lellouch-L\"uscher relation that sheds some light on its relation to spectral functions 

%as well as 

%{\mh %Mention also the Liu papers

%From 1993 Ref.~\cite{Liu:1993cv}

%I think that this is the first one to mention MEM, Ref.~\cite{Liu:2017lpe}

%And the lattice proceedings Ref.~\cite{Liu:2016djw}

%In addition Shoji's B decay paper is Ref.~\cite{Hashimoto:2017wqo}

%And the Adelaide paper mentioned by Shoji is Ref.~\cite{Chambers:2017dov}
%}

\bigskip

The remainder of this paper is organized as follows. In the following section we detail our formalism for estimating widths and differential rates from Euclidean correlators. In Sec.~\ref{sec:examples} we describe two specific examples where our approach may be applied, the transition regime between elastic and deep inelastic scattering as well as semi-leptonic heavy-flavor decays. Next, in Sec.~\ref{sec:LL} we discuss the relation of our approach to the Lellouch-L\"uscher formalism of Ref.~\cite{Lellouch2000}. This is followed by a numerical example of the Backus-Gilbert method applied to a toy system in Sec.~\ref{sec:num}. We close with brief conclusions.

\section{Formalism\label{sec:FGR}}

In this section we explain our approach for estimating total decay and transition rates from lattice QCD. We also discuss how the technique may be used to study photo-production processes and semi-leptonic decays with differential rates in the photon or lepton-neutrino invariant mass squared.

We begin with a strongly-interacting quantum field theory, described by the hamiltonian density $\mathcal H_{\rm QCD}(x)$, and including a stable single-particle state satisfying
\begin{equation}
\left [ \int d^3 \textbf x \, \mathcal H_{\rm QCD}(x)  \right ] \vert N, \textbf P, \lambda \rangle = \vert N, \textbf P, \lambda \rangle E_{N} \,,
\end{equation} 
where $E_{N} = \sqrt{M_N^2 + \textbf P^2}$ and $M_N$ is the physical mass of the particle. Here we have in mind a nucleon state, but our formalism holds for any particle that is stable under the strong interaction. In addition to a flavor label, $N$, and total three-momentum, $\textbf P$, we have included $\lambda$ to denote the azimuthal component of the particle's intrinsic spin.

We next introduce the infinite-volume matrix element
\begin{equation}
\label{eq:TransAmp}
\mathcal A_{N(\lambda) \to \alpha}(E, \textbf p) \equiv \langle E, \textbf p, \alpha; \mathrm{out} \vert  \mathcal J_{\mathcal Q}(0) \vert N, \textbf P, \lambda \rangle \,.
\end{equation}
Here $\mathcal J_{\mathcal Q}(x)$ is a local current and $\langle E, \textbf p, \alpha; \mathrm{out}  \vert$ is a multi-hadron out-state with energy $E$, total momentum $\textbf p$, and all other quantum numbers labeled by the combined index $\alpha$. The multi-particle states have standard relativistic normalization. For example, a two-particle state satisfies
\begin{equation}
\langle  E, \textbf p; N \pi, \textbf k; \mathrm{out} \vert E', \textbf p'; N \pi, \textbf k'; \mathrm{out}\rangle = 2 \omega_{N, \textbf k} 2 \omega_{\pi, \textbf p - \textbf k} (2 \pi)^6 \delta^3(\textbf k - \textbf k') \delta^3(\textbf p - \textbf k - \textbf p' + \textbf k')\,,
\end{equation}
where we have set the collective index $\alpha$ to represent a two-particle state comprised of a pion and a nucleon, with the nucleon carrying momentum $\textbf k$ or $\textbf k'$, and have also defined
\begin{equation}
\omega_{N, \textbf k} = \sqrt{M_N^2 +  \textbf k^2} \,, \ \ \ \ \  \omega_{\pi, \textbf p - \textbf k} = \sqrt{M_\pi^2 + (\textbf p - \textbf k)^2}   \,.
\end{equation}
Throughout this work, we denote the number of hadrons in an asymptotic state by $N_\alpha$, e.g.~in this case $N_\alpha = 2$.

In Eq.~(\ref{eq:TransAmp}) we have allowed the energy and momentum of the final state to differ from the single-hadron initial state. This type of matrix element is appropriate for describing transitions in which photons or leptons, represented by the current $\mathcal J_{\mathcal Q}$, inject or carry away some amount of energy and momentum. Another case of interest is when the current represents an insertion of the weak hamiltonian, mediating a decay into a purely hadronic final state. In this case we denote the initial state by $\vert D, \textbf P \rangle$ and the operator by $\mathcal H_{\mathcal Q}$ and define
\begin{equation}
\mathcal A_{D \to \alpha} \equiv \langle E_D, \textbf P, \alpha; \mathrm{out} \vert  \mathcal H_{\mathcal Q}(0) \vert D, \textbf P \rangle \,.
\end{equation}
Here we have in mind the $D$-meson of the Standard Model, which can decay into many multi-hadron final states. Again the labeling is only suggestive, as the formalism applies to any QCD-stable states.

In lattice QCD it is only possible to calculate finite-volume matrix elements of the form
\begin{align}
\label{eq:FVME}
M_{k, N(\lambda) \to \mathcal Q}(\textbf p, L) & \equiv \langle E_k(L), \textbf p, \mathcal Q \vert  \mathcal J_{\mathcal Q}(0) \vert N, \textbf P, \lambda \rangle_L \,, \\
M_{k, D \to \mathcal Q}(L) & \equiv \langle E_k(L), \textbf P, \mathcal Q \vert  \mathcal H_{\mathcal Q}(0) \vert D, \textbf P \rangle_L \,.
\label{eq:FVME2}
\end{align}
Our convention is such that the state  $\mathcal J_{\mathcal Q}(0) \vert N, \textbf P, \lambda \rangle $ has quantum numbers $\mathcal Q$, and so the final state must have the same quantum numbers, as indicated. We define the finite-volume states with unit normalization throughout, e.g.~$\langle E_k(L), \textbf p, \mathcal Q \vert E_k(L), \textbf p, \mathcal Q \rangle = 1$.

It is non-trivial to relate the finite-volume matrix elements, $M_{k, N(\lambda) \to \mathcal Q}(\textbf p, L)$ and $M_{k, D \to \mathcal Q}(L)$, to the transition amplitudes, $\mathcal A_{N(\lambda) \to \alpha}(E, \textbf p)$ and $\mathcal A_{D \to \alpha}$. As was discussed in the introduction, and has been argued in various contexts in Refs.~\cite{Lellouch2000,KSS2005,Christ2005,MeyerTimelike2011,HSmultiLL,BricenoTwoPart2012,BHWOneToTwo,BHOneToTwoSpin}, the finite-volume matrix element is equal to a linear combination of all infinite-volume matrix elements with asymptotic final states that carry the same quantum numbers. In the finite volume it is not possible to define asymptotic states, and therefore one cannot isolate exclusive decay amplitudes. 
For example, if one considers charm to strange decays, $D \to s + X$, then one must include out-states such as $\overline K \pi$, $\overline K \pi \pi$, $\overline K \pi \pi \pi$, $\overline K \pi \pi \pi \pi$, $\overline K K \overline K$.

\bigskip

In this work we present an alternative approach that allows one to directly calculate transition rates that are integrated over all hadronic kinematics, but are differential rates with respect to non-hadronic degrees of freedom. 
To define such total rates one requires the standard Lorentz-invariant phase-space measure for an $N_\alpha$-particle state 
\begin{equation}
d \Phi_\alpha(k_1, \cdots, k_{N_{\alpha}}) \equiv   \frac{d^3 \textbf k_1}{(2 \pi)^3 2 \omega_{\textbf k_1} } \cdots   \frac{d^3 \textbf k_{N_\alpha}}{(2 \pi)^3 2 \omega_{\textbf k_{N_\alpha}} } (2 \pi)^4 \delta^4(P - {\textstyle\sum_{i=1}^{N_\alpha}} k_i) \,.
\end{equation}
The phase-space measure can be used, for example, to express total decay widths according to 
\begin{equation}
\label{eq:Gammadef}
\Gamma_{D \to \mathcal Q} \equiv \frac{1}{2 M_{D} }  \sum_{ \alpha} \frac{1}{S_\alpha} \int d \Phi_\alpha(k_1, \cdots, k_{N_\alpha}) \vert  \langle  E_{D} , \textbf P , \alpha; \mathrm{out} \vert  \mathcal H_{\mathcal Q}(0)   \vert D, \textbf P \rangle \vert^2  \,,
\end{equation}
where integration runs over all real values of all the three-momenta in the measure. At leading order in the weak interaction, $\Gamma_{D \to \mathcal Q}$ 
gives the total width of the $D$-meson into all open hadronic channels with quantum numbers $\mathcal Q$. 
Note that, in the mother particle's rest frame, this is an alternative version of Eq.~(\ref{eq:FGR}) in which the infinite-volume limit has been rewritten as a phase-space integral.

Both differential and total rates can be directly extracted 
from a more general object that we call the {\em transition spectral function} and define as
\begin{equation}
\label{eq:rhodef}
\rho_{\mathcal Q, \textbf P}(E, \textbf p) \equiv \frac{1}{n_\lambda} \sum_{\lambda, \alpha} \frac{1}{S_\alpha} \int d \Phi_\alpha(k_1, \cdots, k_{N_\alpha}) \vert  \langle  E , \textbf p , \alpha; \mathrm{out} \vert  \mathcal J_{\mathcal Q}(0)   \vert N, \textbf P, \lambda \rangle \vert^2 \,,
\end{equation}
where the sum runs over all states in the Hilbert space carrying the four-momentum $(E, \textbf p)$, and where $S_\alpha$ is the symmetry factor that avoids double counting phase-space points related by the exchange of identical particles. Here we use the notation appropriate for the nucleon. 

To rewrite $\rho_{\mathcal Q, \textbf P}(E, \textbf p)$ in a more useful form, we now apply a Fourier transform, together with its inverse, to reach 
\begin{multline}
  \rho_{\mathcal Q, \textbf P}(E, \textbf p) =  \frac{1}{n_\lambda} \sum_\lambda   \int d^4 x \ e^{ i (E -  E_{N}) t - i (\textbf p - \textbf P) \cdot \textbf x}  \int \frac{d^3 \textbf p'}{(2 \pi)^3 }  \int \frac{dE'}{2 \pi}  e^{- i (E'-E_{N}) t + i (\textbf p'- \textbf P) \cdot \textbf x}  \\ 
    \times    \sum_\alpha \frac{1}{S_\alpha} \int d \Phi_\alpha(k_1, \cdots, k_{N_\alpha}) \langle N, \textbf P, \lambda  \vert  \mathcal J^\dagger_{\mathcal Q}(0) \vert E', \textbf p', \alpha; \mathrm{out}  \rangle \langle  E', \textbf p', \alpha; \mathrm{out}   \vert  \mathcal J_{\mathcal Q}(0)   \vert N, \textbf P, \lambda \rangle \,.
\end{multline}
%For simplicity we have assumed that the current $\mathcal J_{\mathcal Q}$ is self-adjoint.
The expression can be further simplified by moving the second exponential inside the left-most matrix element and identifying it as a standard translation operator on the current
\begin{multline}
\label{eq:xinH}
  \rho_{\mathcal Q, \textbf P}(E, \textbf p) =  \frac{1}{n_\lambda} \sum_\lambda  \int d^4 x \ e^{ i (E -  E_{N}) t - i (\textbf p - \textbf P) \cdot \textbf x}  \int \frac{d^3 \textbf p'}{(2 \pi)^3 }  \int \frac{dE'}{2 \pi}    \\ 
    \times    \sum_\alpha \frac{1}{S_\alpha} \int d \Phi_\alpha(k_1, \cdots, k_{N_\alpha}) \langle N, \textbf P, \lambda  \vert  \mathcal J^\dagger_{\mathcal Q}(x) \vert E', \textbf p', \alpha; \mathrm{out}  \rangle \langle  E', \textbf p', \alpha; \mathrm{out}   \vert  \mathcal J_{\mathcal Q}(0)   \vert N, \textbf P, \lambda \rangle \,,
\end{multline}
where the left-most current is now evaluated at $x = (t, \textbf x)$.

We now identify the integrals over $E'$ and $\textbf p'$ together with the sum over $\alpha$ and integrals over $d \Phi_\alpha$, as a sum over all states with quantum numbers $\mathcal Q$ in the QCD Hilbert space. This sum over states, together with the outer product appearing between the factors of $\mathcal J_{\mathcal Q}$, defines an insertion of the identity. Thus Eq.~(\ref{eq:xinH}) can be rewritten as
\begin{equation}\label{eq:transSF}
\rho_{\mathcal Q, \textbf P}(E , \textbf p)  = \frac{1}{n_\lambda} \sum_\lambda  \int d^4 x \ e^{ i (E -  E_{N}) t - i (\textbf p - \textbf P) \cdot \textbf x} \  \langle N, \textbf P, \lambda  \vert  \mathcal J^\dagger_{\mathcal Q}(x)  \mathcal J_{\mathcal Q}(0)   \vert N, \textbf P, \lambda \rangle \,.
\end{equation}
We deduce that the transition spectral function can be written as the expectation value of a
product of field operators in a one-particle external state. We emphasize here that the matrix element on the right-hand side 
is evaluated in infinite volume, with real Minkowski time coordinates and is not time ordered. 

In  Sec.~\ref{sec:examples}, we describe in specific cases how the transition spectral function can be used to compute 
decay rates and cross-sections. Here we simply note that the total decay width  (\ref{eq:Gammadef}) into hadronic final states 
can be written as 
\begin{equation}
\label{eq:SFtoWidth}
\Gamma_{D \to \mathcal Q} = \frac{1}{2 M_D} \rho_{\mathcal Q, \textbf P}(E_D , \textbf P)  =  \frac{1}{2 M_D}  \int d^4 x  \  \langle D, \textbf P  \vert  \mathcal H_{\mathcal Q}(x)  \mathcal H_{\mathcal Q}(0)   \vert D, \textbf P \rangle  \,.
\end{equation}
The main focus of this work, however, is the more interesting case in which the energy of the outgoing hadrons differs from that of the initial state.

\bigskip

At this point, we have established that the transition spectral function, defined in Eq.~(\ref{eq:rhodef}), gives access to differential transition rates and total decay rates, and have also shown how it may be expressed as a matrix element with two current insertions. However, the discussion thus far, summarized by Eqs.~(\ref{eq:transSF}) and (\ref{eq:SFtoWidth}), has relied crucially on the fact that all quantities are defined in an infinite volume and with real, Minkowski-signature time coordinates. Thus, the relations do not seem to be of relevance for calculations in lattice QCD, necessarily restricted to a finite volume and to Euclidean signature.

To bridge this gap, we now consider the finite-volume Euclidean correlator most closely related to that used above
\begin{equation}
\label{eq:fourpoint}
G_{\mathcal Q, \textbf P}(\tau,\textbf x, L) \equiv 2 E_{N} L^6 e^{- E_{N} \tau + i \textbf P \cdot \textbf x }   \lim_{\tau_f \to \infty}  \lim_{\tau_i \to - \infty}   \frac{ \sum_\lambda \langle \Psi_\lambda(\tau_f, \textbf P)  \mathcal J^\dagger_{\mathcal Q}(\tau, \textbf x)  \mathcal J_{\mathcal Q}(0)   \Psi_\lambda^\dagger(\tau_i , \textbf P)  \rangle_{\rm conn} }{\sum_\lambda \langle \Psi_\lambda(\tau_f, \textbf P)      \Psi_\lambda^\dagger(\tau_i , \textbf P)  \rangle} \,,
\end{equation}
where $\Psi_\lambda^\dagger(\tau_i, \textbf P)$ is an interpolator for the nucleon with total momentum $\textbf P$ and spin component $\lambda$, and the subscript ``conn'' indicates subtraction of $\langle  \mathcal J^\dagger_{\mathcal Q}(\tau, \textbf x)  \mathcal J_{\mathcal Q}(0)  \rangle \langle \Psi_\lambda(\tau_f, \textbf P)  \Psi_\lambda^\dagger(\tau_i , \textbf P)  \rangle$. Throughout this work we take $\tau>0$. Evaluating the large time limits we reach
\begin{equation}
G_{\mathcal Q, \textbf P}(\tau,\textbf x, L)  = 2 E_{N} L^3  e^{- E_{N} \tau + i \textbf P \cdot \textbf x} \frac{1}{n_\lambda} \sum_\lambda \langle N, \textbf P, \lambda \vert       \mathcal J^\dagger_{\mathcal Q}(\tau, \textbf x)  \mathcal J_{\mathcal Q}(0)     \vert N, \textbf P, \lambda \rangle_L  \,.
\end{equation}
Next we project the current to definite three-momentum, defining
\begin{align}
\label{eq:GtildeDef}
\widetilde G_{\mathcal Q, \textbf P}(\tau, \textbf p ,L) & \equiv \int d^3 \textbf x \, e^{-i \textbf p \cdot \textbf x} G_{\mathcal Q, \textbf P}(\tau,\textbf x, L)  \,, \\
& = 2 E_{N} L^3 e^{- E_{N} \tau }  \int d^3 \textbf x \,   e^{ - i (\textbf p - \textbf P ) \cdot \textbf x} \frac{1}{n_\lambda} \sum_\lambda \langle N, \textbf P, \lambda \vert       \mathcal J^\dagger_{\mathcal Q}(\tau, \textbf x)  \mathcal J_{\mathcal Q}(0)     \vert N, \textbf P, \lambda \rangle_L \,.
\label{eq:Gtilde2}
\end{align}
Finally, inserting a complete set of finite-volume states into Eq.~(\ref{eq:Gtilde2}) gives
\begin{align}
\label{eq:GtildeCompset}
\widetilde G_{\mathcal Q, \textbf P}(\tau, \textbf p ,L) & = 2 E_{N} L^6 \sum_k e^{- E_k(L) \tau }   \vert M_{k, D \to \mathcal Q}(\textbf p, L) \vert^2  \,,
\end{align}
where the squared magnitudes on the right-side are given by spin averaging the squared magnitudes of the matrix elements defined in Eq.~(\ref{eq:FVME}).

\bigskip

Equation~(\ref{eq:GtildeCompset}) can be rewritten as
\begin{equation}
\label{eq:Lap}
\widetilde G_{\mathcal Q, \textbf P}(\tau, \textbf p ,L)  =  \int_0^\infty \frac{ d \omega}{2 \pi} \,  e^{-\omega \tau}    \rho_{\mathcal Q, \textbf P}(\omega, \textbf p, L) \,,
\end{equation}
where
\begin{equation}
\label{eq:FVSE}
\rho_{\mathcal Q, \textbf P}(E, \textbf p, L) \equiv 2 E_{N} \,L^6 \sum_{k} \vert  M_{k, N \to \mathcal Q} (\textbf p, L)    \vert^2\;  2 \pi\,   \delta \big( E - E_k(L) \big ) \,,
\end{equation}
is the finite-volume spectral function. Substituting this sum of delta functions into Eq.~(\ref{eq:Lap}) and evaluating the integral immediately gives back Eq.~(\ref{eq:GtildeCompset}). We emphasize at this stage that, while the infinite-volume spectral function gives direct access to the decay width, in the finite volume we have a sum of delta peaks. Naively sampling $\rho_{\mathcal Q, \textbf P}(E, \textbf p, L)$ at a specific energy cannot give any useful information as the result will either vanish or diverge.

In order to recover the total decay width, we need to construct a sensible infinite-volume limit of the spectral function. To do so we introduce $\widehat \delta_{\Delta}(\overline \omega, \omega)$ as a regularized delta function, centered at $\overline \omega$ with width $\Delta$. We require only that this satisfies
\begin{equation}
\label{eq:area}
\int_{0}^\infty d\omega\, \widehat \delta_\Delta(\overline \omega,\omega) = 1, \qquad 
\lim_{\Delta\to0}\int_0^\infty d\omega\,\widehat\delta_\Delta(\overline \omega,\omega) \phi(\omega) = \phi(\overline \omega) \,,
\end{equation}
for a smooth test function $\phi(\omega)$. In our approach, $\delta_\Delta(\overline \omega,\omega)$ will
tend to zero exponentially for large $\omega$. We then define
\begin{equation}
\label{eq:smoothSF}
\widehat \rho_{\mathcal Q, \textbf P}(\overline \omega , \textbf p, L, \Delta)  \equiv \int_0^\infty  d \omega\;  \widehat \delta_\Delta(\overline \omega ,\omega)\,  \rho_{\mathcal Q, \textbf P}( \omega , \textbf p, L)  \,.
\end{equation}
This smoothing procedure replaces the sum over delta functions with a smooth function that has a well-defined infinite-volume limit. 

In particular the smoothing allows us to make contact with the discussion of Fermi's Golden Rule in the introduction. There we described how the width of the mother particle is given by studying the system in finite volume, and summing over the squared magnitudes of individual finite-volume matrix elements, multiplied by regularized delta functions. That construction is identical to that of the smoothed, finite-volume spectral function, defined in Eq.~(\ref{eq:smoothSF}). Thus it follows that differential transition rates can be accessed
from the lattice framework via the limits
\begin{equation}
\rho_{\mathcal Q, \textbf P}(E, \textbf p) = \lim_{\Delta \to 0} \lim_{L \to \infty}    \widehat \rho_{\mathcal Q, \textbf P}(E , \textbf p, L, \Delta) \,,
\end{equation}
where the order of limits is important. 

\bigskip

The Backus-Gilbert method applied to the inverse problem of determining $\rho_{\mathcal Q, \textbf P}(E, \textbf p, L)$ 
from $\widetilde G_{\mathcal Q, \textbf P}(\tau, \textbf p, L)$
leads precisely to smoothed quantities of the form given in Eq.~(\ref{eq:smoothSF}). 
If the correlation function is known at a discrete set of Euclidean times, $\tau_j$, then the `resolution function' $\widehat\delta_\Delta(\bar\omega,\omega)$ should be constructed from the Laplace kernel,
\begin{equation}\label{eq:resfct}
\widehat\delta_\Delta(\overline \omega,\omega)  = \sum_{j} C_j(\overline \omega, \Delta) \; e^{-\omega\tau_j},
\end{equation}
with coefficients $C_j$ chosen so as to minimize the width 
\begin{equation}
\label{eq:mincon}
\Delta = \int_0^\infty d\omega \, (\bar\omega-\omega)^2 \, \widehat \delta_{\Delta}(\overline \omega,\omega)^2  \,,
\end{equation}
under the unit-area constraint of Eq.~(\ref{eq:area}). The Backus-Gilbert method then yields an estimate of the smoothed spectral function
\begin{equation}
\label{eq:rhohatest}
\widehat \rho_{\mathcal Q, \textbf P}(\overline \omega , \textbf p, L, \Delta) =2 \pi \sum_j C_j(\overline \omega, \Delta) \; \widetilde G_{\mathcal Q, \textbf P}(\tau_j, \textbf p, L) \,.
\end{equation}

\begin{figure}
\begin{center}
\includegraphics[scale=0.7]{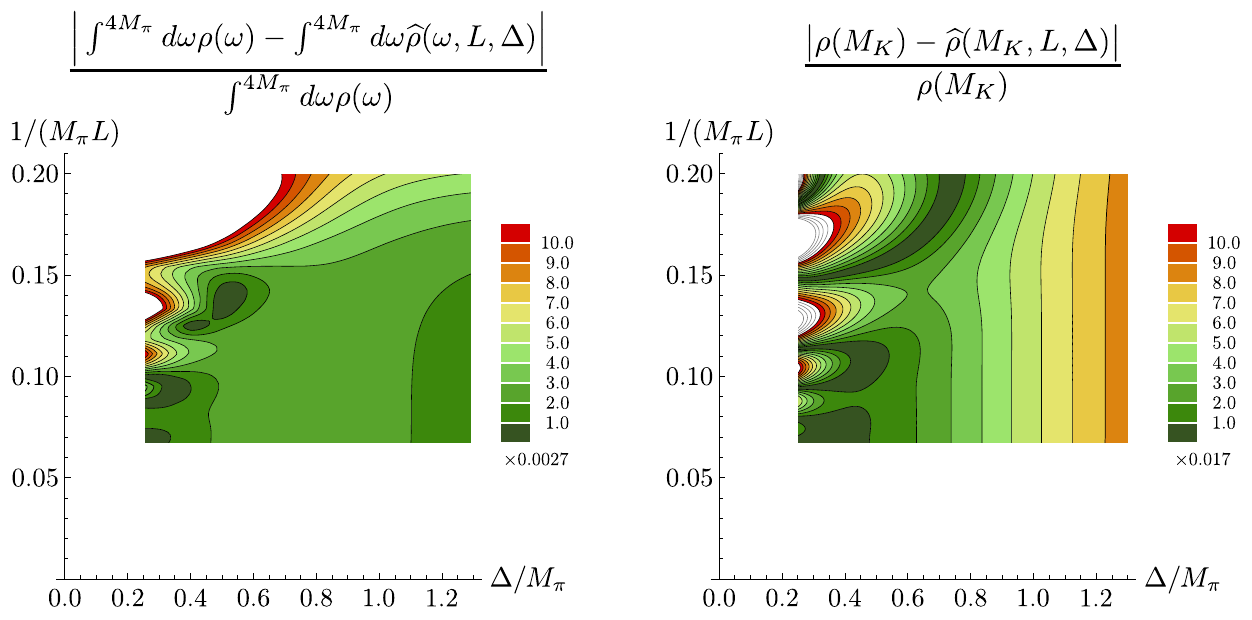}
\caption{Accuracy of the infinite-volume spectral function as estimated from the smeared finite-volume spectral function for various values of $M_\pi L$ and $\Delta/M_\pi$. These plots were constructed assuming a single channel of non-interacting pions coupled via a constant amplitude $A_{K \to \pi \pi}$ to an initial kaon. The smearing here was performed using a normalized Gaussian for the resolution function $\widehat \delta_\Delta$ with $\Delta$, defined in Footnote 4, differing from the definition of Eq.~(\ref{eq:mincon}) by a simple numerical factor. In the left panel we plot the difference of the spectral function integrated from $E=0$ to $E=4 M_\pi$, whereas in the right panel we consider the fixed value $E=M_K$, that directly gives the decay rate.  \label{fig:optimal}}
\end{center}
\end{figure}

It is well-known that, due to the numerically ill-posed nature of the
inverse Laplace transform, reducing the width $\Delta$ is
computationally very costly.  In the case of the $K\to\pi\pi$ decay,
 this method is therefore likely to be less accurate than the
Lellouch-L\"uscher method. Nevertheless, we note that the positivity property of the spectral
function ameliorates the inverse problem. To gain some intuition on this, in Fig.~\ref{fig:optimal} we show the accuracy of estimating the infinite-volume spectral function via the smeared finite-volume result, for various values of $1/(M_\pi L)$ and $\Delta/M_\pi$, in the case of a constant decay amplitude $A_{K \to \pi \pi}$ and negligible interactions of the outgoing pions. In this example we use a normalized Gaussian for the resolution function, $\widehat \delta_{\Delta}$.%
%%%
%%%
%%%
\footnote{
More precisely we define
\begin{equation}
\widehat \delta_{\Delta}(\overline \omega, \omega) = \frac{1}{\sqrt{2 \pi} \Delta} e^{-(\overline \omega - \omega)^2/(2\Delta^2)} \,.
\end{equation}
We caution that the $\Delta$ used here differs from that defined in Eq.~(\ref{eq:mincon}) by a factor of $4 \sqrt{\pi}$.
}
%%%
%%%
%%%
While not equivalent to the resolution functions that result from the Backus-Gilbert method, these results give an idea of the optimal trajectory to follow in the $[ \Delta/M_\pi, 1/(M_\pi L) ]$ plane.  In particular it is manifest that, if one decreases $\Delta/M_\pi$ at a fixed $M_\pi L$ then the estimator completely fails at some stage, when the value of width becomes too small. This emphasizes the importance of the order in which the limits are to be taken.

While unlikely to be competitive with the Lellouch-L\"uscher approach for a single two-particle channel, our method has
the advantage of not requiring a detailed understanding of the
connection between the finite-volume and infinite-volume matrix
elements, which presumably becomes untractable when many channels are
open. As already mentioned in the introduction, for differential 
transition rates the limited energy resolution means that 
our approach can only yield predictions in relatively broad `energy bins' of width $\Delta$,
but it does not imply an uncontrolled systematic error, given that 
the resolution function (\ref{eq:resfct}) is known exactly. 
In the following section we discuss various specific examples
where the formalism presented here seems particularly promising.

\section{Example applications \label{sec:examples}}

In this section we present two examples of phenomenologically interesting cases where our formalism for extracting total transition rates might be applied. In Sec.~\ref{sec:dis}, we discuss deep inelastic scattering, and in Sec.~\ref{sec:semilep}, semi-leptonic decays of heavy hadrons.

\subsection{Deep inelastic scattering\label{sec:dis}}

Deep inelastic scattering (DIS), the collision of high-energy leptons with hadrons via virtual photon exchange, has played a major role in particle and nuclear physics. The deep inelastic scattering experiments in the late 1960's revealed structure within the nucleon with initially surprising scaling laws, leading to partonic models and eventually to the formulation of quark and gluon degrees of freedom in QCD. In this sub-section we describe the application of our general formalism to studying deep inelastic scattering in lattice QCD.

\begin{figure}
\begin{center}
\includegraphics[width=0.4\textwidth]{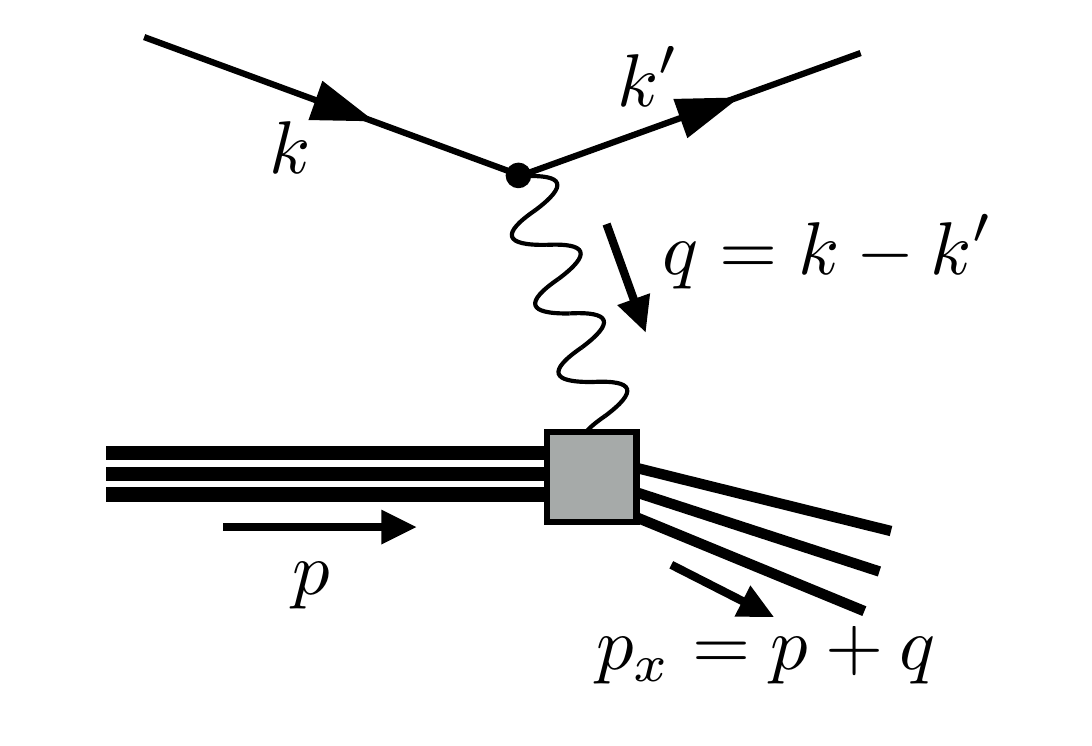}
\vspace{-10pt}
\caption{The kinematics of deep inelastic scattering. A proton with momentum $p$ collides with a hard lepton carrying momentum $k$ via exchange of a virtual, space-like photon with momentum $q$. \label{fig:DIS}}
\end{center}
\end{figure}

To make the presentation self-contained, we review some of the basics of DIS. In Fig.~\ref{fig:DIS} we give a schematic of the events of interest. A hadron, usually a nucleon, with mass $M$ and four-momentum $p$ collides with a hard lepton carrying momentum $k$ via the exchange of a virtual, space-like photon with momentum $q$. Defining $k'$ as the outgoing lepton momentum, we have $q \equiv  k - k'$. The final hadronic state produced in the collision is not detected, so that we only have information about the total rates into all allowed states with momentum $p_x = p + q = p + k - k'$. It is further convenient to define $Q^2 = - q^2$, which is positive for space-like photon momenta. We also introduce the Lorentz-invariant parameters
\begin{equation}
\label{eq:xnudef}
\nu = \frac{q \cdot p}{M}\,,  \ \ \ \  x = \frac{Q^2}{2 M \nu}\,.
\end{equation}

Following the discussion of Ref.~\cite{Manohar1992}, we note that the unpolarized cross-section in the nucleon rest-frame
\begin{equation}\label{eq:DISxsec}
d \sigma = \frac{e^4}{Q^4}   \int \frac{d^3 \textbf k'}{(2 \pi^3) 2 \omega_{\textbf k'} }    \frac{4 \pi \,\ell^{\mu \nu}\,  W_{\mu \nu} (p, k - k')}{2 k^0  \; 2 M \,(v_{\rm rel} = 1)} \,,
\end{equation}
is given by contracting the hadronic tensor, $W_{\mu \nu}$, discussed in the following paragraphs, with the leptonic tensor, $\ell^{\mu \nu}$, defined as
\begin{align}
\ell^{\mu \nu}  & = \sum_{s'_{\ell}}    \overline u_{s'_\ell}(k')     \gamma^\nu  u_{s_\ell}(k) \bar u_{s_\ell}(k)     \gamma^\mu            u_{s'_\ell}(k')  \,, \\
& = 2 \big ( k^\mu k'^\nu + k^\nu k'^\mu - g^{\mu \nu} k \cdot k' - i \epsilon^{\mu \nu \alpha \beta} q_\alpha s_{\ell \beta}   \big) + \mathcal O(m_\ell/\Lambda) \,,
\end{align}
where $m_\ell$ is the lepton mass and $\Lambda$ a typical scale of the scattering process. Here $u_s$ and $\bar u_s$ are standard spinors, projected to spin state $s$ and normalized so that
$ \sum_{s}    u_{s}(k)   \overline u_{s}(k)  = \slashed k + m_\ell \,$,    
and the $\gamma^\mu$ are standard Dirac matrices. Note that the spin-independent part of $\ell^{\mu \nu}$ is symmetric, while the spin-dependent part is anti-symmetric. Thus an unpolarized lepton beam only probes the symmetric part of the hadronic tensor and the polarization asymmetry of cross sections depends only on the hadronic tensor's anti-symmetric structure. Here we focus on the unpolarized case.

The spin-averaged hadronic tensor, denoted $W_{\mu \nu}$, is defined as
\begin{equation}
W_{\mu \nu }(p,q) = \frac{1}{4 \pi\,n_\lambda} \sum_{\lambda} \int d^4 x \, e^{i q \cdot x} \, \langle N, \textbf p, \lambda \vert  j_\mu(x) j_\nu(0)  \vert N, \textbf p, \lambda \rangle  \,,
\end{equation}
where $n_\lambda=2$ for the nucleon and $j_\mu=\sum_f Q_f \overline \psi_f \gamma_\mu \psi_f$ is the electromagnetic current, expressed as a sum over all flavors $f$ of quark bilinears, weighted by their charge, $Q_f$. The notation here is slightly different from the previous section with the incoming nucleon carrying three-momentum $\textbf p$.
Following Ref.~\cite{Manohar1992}, we write
\begin{equation}
W_{\mu\nu} = F_1\left(-g_{\mu\nu}+ \frac{q_\mu q_\nu}{q^2}\right) 
+ \frac{F_2}{p\cdot q} \left(p_\mu - \frac{p\cdot q\;q_\mu}{q^2}\right) \left(p_\nu - \frac{p\cdot q\;q_\nu}{q^2}\right) \,,
\end{equation}
where $F_1$ and $F_2$ are the structure functions and depend only on $q^2$ and $p \cdot q$. Defining 
$s_{pq}= M^2  -{(p\cdot q)^2}/{q^2}  \,$,
the individual structure functions can be isolated by taking the linear combinations
\begin{align}
F_1 &= \frac{1}{2}\Big[- W^\mu{}_\mu + \frac{1}{s_{pq}} \;p^\mu p^\nu W_{\mu\nu} \Big] \,, \\
F_2 &= \frac{p\cdot q}{2s_{pq}}
\Big[-W^\mu{}_\mu + \frac{3}{s_{pq}}\;p^\mu p^\nu W_{\mu\nu}\Big] \,.
\end{align}
The unpolarized cross-section (\ref{eq:DISxsec}) for the inclusive $e+p\to e'+{\rm hadrons}$ process can be written
in terms of these as
\begin{equation}
\frac{d^2\sigma}{dxdy} = \frac{e^4ME}{2\pi Q^4}\Big[xy^2 F_1+(1-y)F_2 \Big],
\end{equation}
where $y=(p\cdot q)/(p\cdot k)$ represents the fractional energy loss of the lepton
in the nucleon rest frame.

\bigskip

Combining the definition of $W_{\mu\nu}$ with the discussion of the previous section, the task is to invert 
\begin{equation}
\widetilde G_{\mu \nu, \textbf p}(\tau,\textbf p_x, L) = \frac{1}{2 \pi} \int_0^\infty d \omega \,  e^{-\omega \tau}   \rho_{\mu \nu, \textbf p} \big (\omega, \textbf p_x  , L\big )  \,,
\end{equation}
with the correlator
\begin{equation}\label{eq:G_DIS}
\widetilde G_{\mu \nu, \textbf p}(\tau,\textbf p_x, L) \equiv 2 E_{\textbf p} L^6 e^{- E_{\textbf p} \tau}   \int d^3 \textbf x \, e^{- i  \textbf q  \cdot \textbf x}    \lim_{\tau_f \to \infty}  \lim_{\tau_i \to - \infty}   \frac{ \sum_\lambda \langle \Psi_\lambda(\tau_f, \textbf p)   j_{\mu}(\tau, \textbf x)  j_{\nu}(0)   \Psi_\lambda^\dagger(\tau_i , \textbf p)  \rangle_{\rm conn} }{\sum_\lambda \langle \Psi_\lambda(\tau_f, \textbf p)      \Psi_\lambda^\dagger(\tau_i , \textbf p)  \rangle} \,,
\end{equation}
where $E_{\textbf p} = \sqrt{M^2 + \textbf p^2}$. The important point is that the role of $\omega$, in terms of which the 
spectral representation is written, is played by the energy $p_x^0$ of the final state.

Applying the Backus-Gilbert method leads to a smoothed spectral function
\begin{equation}
\widehat \rho_{\mu \nu, \textbf p} \big (p_x^0, \textbf p_x , L, \Delta \big ) \equiv    \int_{0}^\infty d \omega  \, \widehat \delta_{\Delta}( p_x^0, \omega) \, \rho_{\mu \nu, \textbf p} \big (\omega, \textbf p_x , L\big ) \,,
\end{equation}
 from which one can estimate the hadronic tensor via
\begin{equation}
W_{\mu \nu }(p,q)= \frac{1}{4 \pi} \lim_{\Delta \to 0} \lim_{L \to \infty}  \widehat \rho_{\mu \nu, \textbf p} \big (p_x^0, \textbf p_x , L, \Delta \big ) \,. 
\end{equation}
This expression makes manifest how the lattice calculation yields a
sharply defined outgoing total hadron momentum $\textbf p_x$, but only
limited resolution in the corresponding energy $p_x^0$.
The Lorentz-scalar combinations 
\begin{align}
{p^\mu p^\nu } W_{\mu\nu} &= {s_{pq}} \Big[- F_1 +\frac{ s_{pq}}{p\cdot q} F_2 \Big] \,,
\\
W^\mu{}_\mu &= -3F_1 + \frac{ s_{pq}}{p\cdot q} F_2 \,
\end{align}
are the more primary quantities  in our lattice approach than $F_1$ and $F_2$, 
because the kinematic factors $s_{pq}$ and $p\cdot q$ (but not $p^\mu$) are affected by the limited resolution of $p_x^0$.

\begin{figure}
\begin{center}
\includegraphics[width=0.85\textwidth]{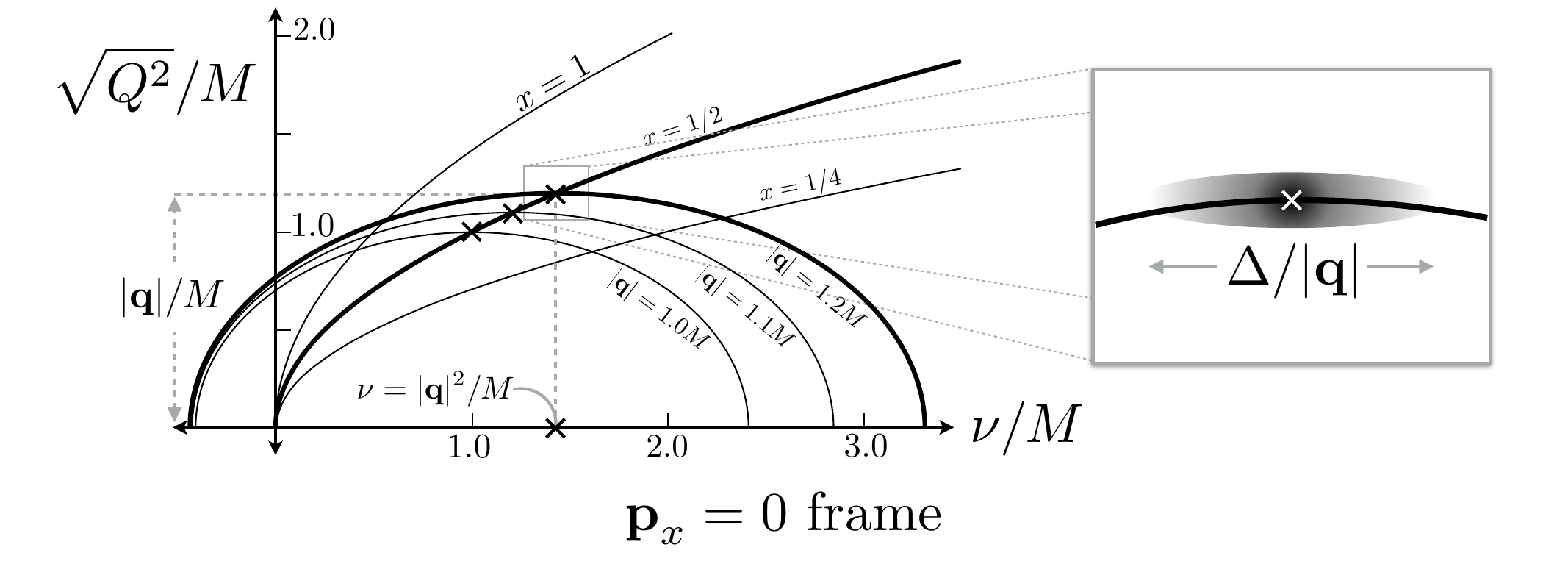}
\caption{Plot of constant-$\vert \textbf q \vert$ contours (ellipses) and constant $x$-contours (square-root functions) in the $\nu, \sqrt{Q^2}$ plane. The smearing limits resolution along the ellipses, as indicated in the right panel. \label{fig:DISkin2}}
\end{center}
\end{figure}

In a lattice calculation, the four variables\footnote{Here we ignore possible violations of O(3)
  rotation symmetry by the cubic box.} $\textbf p^2, \textbf
q^2, \textbf p\cdot \textbf q$ and $p_x^0$ can be varied independently,
with the first three sharply defined, up to statistical 
uncertainties.
Due to Lorentz symmetry however, the nucleon structure functions only depend
on two invariants: the photon virtuality $Q^2$ and the photon energy in
the nucleon rest frame, $\nu$, or alternatively for the latter, the Bjorken variable
 $x$ [see Eq.~(\ref{eq:xnudef})].  
In general, there is therefore a two-parameter family
of lattice kinematic variables that realize a given $(Q^2,x)$ pair.
This redundancy can be exploited to minimize the impact of the limited resolution 
in $p_x^0$. For instance, in the deep inelastic regime, the structure functions
depend only weakly on $Q^2$, and therefore one might wish to 
reduce the impact of the limited resolution in $p_x^0$ on the Bjorken variable $x$
and accept a broader `bin' in $Q^2$.

Explicitly, the connection between the four lattice variables and the relativistic invariant variables reads
\begin{eqnarray}\label{eq:Mnu}
M\nu &=& E_{\textbf p} p_x^0 - E_{\textbf p}^2 - \textbf q \cdot \textbf p \,,
\\
Q^2 &=& \textbf q^2 - (p_x^0-E_{\textbf p})^2.
\label{eq:Qsq}
\end{eqnarray}
At fixed $\textbf p$ and $\textbf q$, $\nu$ is thus an affine function of the dispersion variable $p_x^0$.
Given these relations, it is straightforward to compute the linear propagation of the uncertainty in the variable
$p_x^0$ on the variables $\nu$, $Q^2$ and $x$.
Eliminating the `fuzzy' variable $p_x^0$ from Eqs.\ (\ref{eq:Mnu}--\ref{eq:Qsq}), we obtain 
\begin{equation}
\label{eq:genellipse}
1 = \frac{Q^2}{\textbf q^2} + \frac{M^2}{E_{\textbf p}^2 \textbf q^2} \left( \nu + \frac{\textbf q \cdot \textbf p}{M}  \right )^2 \,.
\end{equation}
In other words, fixed values of $\textbf p$ and $\textbf q$ define an ellipse in the $(\nu, \sqrt{Q^2})$ plane, centered at $(- \textbf q \cdot \textbf p/M, 0)$ with long radius $E_{\textbf p} \vert \textbf q \vert/M$ along $\nu$ and short radius $\vert \textbf q \vert$ along $\sqrt{Q^2}$.
Thus, while the ellipse is `certain' in a given kinematic setup on the lattice, 
the finite resolution in the variable $p_x^0$ results in a spread along the ellipse.
Contours of constant $x$ in the $(\nu, \sqrt{Q^2})$ plane correspond to square-root functions. 

Of particular interest is the point of maximum virtuality $Q^2$ 
for given $\textbf q$, which is reached when $p_x^0=E_{\textbf p}$, implying 
\begin{equation}
Q^2=\textbf q^2,\qquad x = \frac{\textbf q^2}{-2\textbf p\cdot \textbf q}.
\end{equation}
At that point, corresponding to the top of the ellipse,
to linear order the uncertainty in $p_x^0$ only affects the variable $\nu$ and not $Q^2$;
at the same level of approximation, the spread $\Delta x$ in Bjorken $x$ reads
\begin{equation}
\frac{\Delta x}{x} = - \frac{\Delta \nu}{\nu} = \frac{E_{\textbf p}^2}{-\textbf p\cdot \textbf q}\;\frac{\Delta}{p_x^0}
\qquad (\textrm{at~}Q^2=\textbf q^2).
\end{equation}

In Fig.~\ref{fig:DISkin2}, we plot the ellipses (\ref{eq:genellipse}) for the case $\textbf p_x =  \textbf p+ \textbf q=  \textbf 0$,
corresponding to the rest frame of the outgoing hadronic state, along with the $x=\,$constant curves. 
In this frame, the maximum value of $Q^2$ at fixed $\textbf q^2$ corresponds to $x=1/2$.
For a more general frame defined by $\textbf p = - \beta \textbf q$,
the maximum value of $Q^2$  corresponds to
\begin{equation}
x = \frac{1}{2 \beta} \,, \qquad \frac{\Delta x}{x} \stackrel{\textbf p^2\gg M^2}{\approx} \beta\, \frac{\Delta}{p_x^0}.
\end{equation}
For $\beta=1/2$, the maximum $Q^2$ point then corresponds to probing elastic $e, p$ scattering in the Breit frame,
$\textbf p = - \textbf q/2 = - \textbf p_x$. For $\beta=1$, $\textbf p_x=0$ and one is probing $x=1/2$.
To reach ever smaller values of $x$ while keeping the virtuality large, $\beta$ must be increased further
and one approaches the `infinite-momentum' frame of the nucleon. However, it becomes increasingly difficult
to maintain a useful resolution in $x$, since the relative spread of $p_x^0$ gets amplified by a factor $\beta$.
Hence $x$ and its spread $\Delta x$ become comparable when $\beta={\mathcal O}(p_x^0/\Delta)$, 
and for a given relative resolution on $p_x^0$, effectively the smallest $x$ that can be probed is $x={\mathcal O}(\Delta/p_x^0)$.

In summary, while there are kinematic limitations and numerical challenges to a lattice
calculation of hadronic structure functions, there does not seem to be
a conceptual obstacle to probing structure functions down to $x\approx
1/3$ at ${Q^2}\approx 4{\rm GeV}^2$. It is well-known to be difficult
to achieve a good signal-to-noise ratio even in nucleon two-point
functions at large separations. This problem is an algorithmic one.
To test the ideas presented here it would be interesting to first
calculate the structure functions of the pion, for which the
signal-to-noise ratio is more favorable, although it also deteriorates
as the pion momentum $\textbf p$ increases. For that reason,
the most promising opportunity for lattice calculations may be 
to cover the transition region from real photons to the onset of 
deep inelastic scattering. The formalism presented may also prove useful
at the conceptual level, e.g.\
in rederiving the theoretical predictions for deep inelastic scattering,
in particular the development of the amplitude in a series of twist-two 
operators or its representation through light-like Wilson lines.
The formalism is also flexible enough to accomodate the spin dependent
structure functions $g_1$ and $g_2$, different local currents, for instance
those determining neutrino DIS or the interference of $\gamma$ and $Z$ exchange, and 
off-forward nucleon matrix elements.

We also note that our method is complementary to the approach of Ji, presented in Ref.~\cite{Ji:2013dva,Chambers:2017dov}. In that work the author shows how structure functions may be studied via equal-time matrix elements in the large momentum limit. Such equal-time matrix elements can be directly extracted from Euclidean correlation functions without solving the inverse problem, but the large momentum limit is challenging in a realistic lattice calculation and the renormalization of the relevant operators is not yet well understood. It would be interesting to compare the two approaches in more detail.

Finally, we remark that the Euclidean correlation function (\ref{eq:G_DIS}),
after applying Euclidean-time-ordering and Fourier-transforming with an imaginary frequency ($e^{\omega\tau}$),
gives direct access~\cite{Ji:2001wha} to the nucleon forward Compton amplitude for photon virtuality $Q^2=\textbf q^2-\omega^2$, 
a very worthwhile goal in itself.

%gives direct access [43,Adelaide_paper] to the nucleon forward
%Compton amplitude

\subsection{Semileptonic decays: $H_Q \to  \ell + \bar \nu_{\ell} + X$\label{sec:semilep}}

\begin{figure}
\begin{center}
\includegraphics[width=0.4\textwidth]{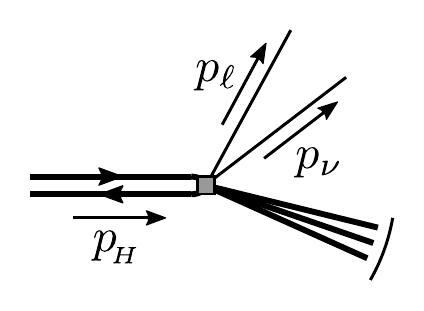}
\vspace{-10pt}
\caption{The kinematics of semi-leptonic decay. The incoming hadron carries momentum $p_H$ and the outgoing lepton and neutrino carry $p_\ell$ and $p_\nu$ respectively. We also define $q = p_\ell + p_\nu$. \label{fig:semilep}}
\end{center}
\end{figure}

The matrix elements relevant for deep inelastic scattering, discussed in the previous subsection, have a character similar to those that enter semi-leptonic weak decays of heavy mesons. The process of interest is shown in Fig.~\ref{fig:semilep}: an incoming hadron, denoted $H_Q$ with four-momentum $p_H$, decays into a lepton, carrying momentum $p_\ell$, an anti-neutrino, with momentum $p_\nu$, and a hadronic state. The subscript $Q$ indicates the heavy quark contained within the incoming hadron. We have in mind mesons with $Q = c$ or $b$, but the approach described here can be used to describe any semi-leptonic decay.

Defining $q \equiv p_\ell + p _\nu$,  the hadronic matrix element needed for $H_Q \to  \ell + \bar \nu_{\ell} + X$ is very similar to that described in the previous subsection: the only distinctions are that $q$ flows away from the vertex and is time-like, and that the current mediating the decay is modified. Following Ref.~\cite{BlokSemiLep1994}, we aim to calculate the differential decay rate with respect to the lepton energy, $E_\ell$, the lepton-neutrino invariant mass, $q^2$, and the lepton-neutrino combined energy, $q^0$. The final result can be expressed as
\begin{equation}
\label{eq:decayrate}
\frac{d^3 \Gamma}{ d E_e d q^2 d q^0} = \vert V_{qQ} \vert^2 \frac{G_F^2}{32 \pi^2} \Big [ 2 q^2 w_1 + [4 E_e (q^0 - E_e) - q^2] w_2 + 2 q^2 (2 E_e - q^0) w_3 \Big]  \,,
\end{equation}
where $V_{qQ}$ is the Cabibbo-Kobayashi-Maskawa (CKM) matrix element describing the flavor change in the decay, $G_F$ is the Fermi decay constant and the $w_i$ are hadronic structure functions.

The structure functions result from decomposing
\begin{equation}
W^{H_Q \to X}_{\mu \nu }(v,q) = \frac{1}{2 M_{H_Q}}   \int d^4 x \, e^{- i q \cdot x} \, \langle H_Q, \textbf p \vert  \mathcal J^\dagger_\mu(x) \mathcal J_\nu(0)  \vert H_Q, \textbf p \rangle  \,,
\end{equation}
where $\mathcal J_\mu= \overline q \gamma_\mu (1 - \gamma_5) Q $ is the flavor-changing current built from one heavy  and one light flavor and $M_{H_Q}$ is the mass of the mother particle. Here we have expressed the tensor in terms of the lepton-neutrino four-momentum, $q^\mu$, together with the four-velocity of the incoming hadron, $v^\mu = p^\mu_{H}/M_{H_Q}$. Given that the tensor only depends on these two four-vectors, one can show that it must satisfy
\begin{equation}
W^{H_Q \to X}_{\mu \nu }(v,q) = - w_1 g_{\mu \nu} + w_2 v_\mu v_\nu - i w_3 \epsilon_{\mu \nu  \alpha \beta} v^\alpha q^\beta + w_4 q_\mu q_\nu + w_5 (q_\mu v_\nu +   v_\mu  q_\nu ) \,.
\end{equation}
This defines the structure functions appearing in Eq.~(\ref{eq:decayrate}).

Thus, the task is to invert 
\begin{equation}
\widetilde G^{H_Q \to X}_{\mu \nu, \textbf p}(\tau,\textbf p_x, L) = \frac{1}{2 \pi} \int_0^\infty d \omega \,  e^{-\omega \tau}   \rho^{H_Q \to X}_{\mu \nu, \textbf p} \big (\omega, \textbf p_x  , L\big )  \,,
\end{equation}
with the correlator
\begin{equation}
\widetilde G^{H_Q \to X}_{\mu \nu, \textbf p}(\tau,\textbf p_x, L) \equiv 2 E_{\textbf p} L^6 e^{- E_{\textbf p} \tau}   \int d^3 \textbf x \, e^{ i  \textbf q  \cdot \textbf x}    \lim_{\tau_f \to \infty}  \lim_{\tau_i \to - \infty}   \frac{  \langle \Psi_Q(\tau_f, \textbf p)   \mathcal J^\dagger_{\mu}(\tau, \textbf x)  \mathcal J_{\nu}(0)   \Psi_Q^\dagger(\tau_i , \textbf p)  \rangle_{\rm conn} }{  \langle \Psi_Q(\tau_f, \textbf p)      \Psi_Q^\dagger(\tau_i , \textbf p)  \rangle} \,,
\end{equation}
where $E_{\textbf p} = \sqrt{M_{H_Q}^2 + \textbf p^2}$. Applying the Backus-Gilbert method leads to a smoothed spectral function from which one can estimate the hadronic tensor via
\begin{equation}
W^{H_Q \to X}_{\mu \nu }(p,q)= \frac{1}{2 M_{H_Q}} \lim_{\Delta \to 0} \lim_{L \to \infty}  \int_{0}^\infty d \omega  \, \widehat \delta_{\Delta}( p_x^0, \omega) \, \rho^{H_Q \to X}_{\mu \nu, \textbf p} \big (\omega, \textbf p_x , L\big ) \,.
\end{equation}

\bigskip

We close this section by commenting that our formalism may also be applied to extract total decay widths for purely hadronic decays. One example where total decay widths are of interest is charm to strange decays, in which a weak vector boson mediates a flavor change, $c\to su \bar d$. This transition, relevant for Cabibbo-allowed $D$-meson decays, is given at leading order by a weak hamiltonian of the form~\cite{Buchalla1996}
\begin{equation}\label{eq:calHQ}
\mathcal H_{\mathcal Q}(x) =  \frac{G_F}{\sqrt{2}} V_{cs}^* V_{ud} \big [\bar s(x) \gamma_\mu (1 - \gamma_5) c(x) \big ] \big [\bar u(x) \gamma_\mu (1 - \gamma_5) d(x) \big] \,,
\end{equation}
where $\bar s(x)$, $c(x)$, $\bar u(x)$, $d(x)$ are Dirac fields in the various flavors.

This weak hamiltonian couples the $D$-meson to many decay channels with two or more hadrons. To study this transition via the Lellouch-L\"uscher approach, one would first need to generalize the formalism to accommodate all multi-hadron states, including those with four outgoing particles. One would then aim to disentangle the individual decay rates by calculating finite-volume matrix elements in different volumes, all tuned so that the final finite-volume state has the same energy as the incoming $D$-meson.  In our approach the total width, $\Gamma_{D \to s + X}$, can be accessed by calculating the appropriate four-point function and then estimating the corresponding spectral function. 

A closely related application is the calculation of the lifetime of
charmed baryons. Providing estimates of the lifetime of the doubly
charmed baryons $\Xi_{cc}^+$~\cite{Mattson:2002vu} and $\Xi_{cc}^{++}$
would help their search at LHCb. The spectroscopy of these heavy
baryons has recently been studied on the
lattice~\cite{Padmanath:2015jea}.

We comment that  operator (\ref{eq:calHQ}) mixes under renormalization with a second operator in which color indices are contracted between the two bilinears. Thus, the full calculation requires defining $\mathcal H_{\mathcal Q}$ as a linear combination with appropriate Wilson coefficients. We direct the reader to Ref.~\cite{Buchalla1996} for details.

\section{Comparison to the Lellouch-L\"uscher method\label{sec:LL}}

In this section we study the relation between the formalism presented above and the method of Lellouch and L\"uscher, described in Ref.~\cite{Lellouch2000}. For the case of $K \to \pi \pi$ decays, Lellouch and L\"uscher derived a relation between finite- and infinite-volume matrix elements
\begin{equation}
\label{eq:LL}
\vert M_{k, K \to \pi \pi} (L) \vert^2  =  \frac{\mathcal C_k}{4   M_K E_k(L)^2 L^9} \  \big \vert A_{K \to \pi \pi} [E_k(L)]  \big \vert^2 \,,
\end{equation}
where
\begin{equation}
\mathcal C_k \equiv  \left(\frac{1}{4 \pi^2 q^2}   \frac{\partial \phi(q)}{\partial q} +   \frac{2 \pi}{  p^2 L^3}  \frac{\partial \delta_{\pi \pi}(p)}{\partial p}   \right )_{q= Lp/(2\pi),\ \ p = \sqrt{E_k(L)^2/4-M_\pi^2}}^{-1}  \,.
\end{equation}
Here $\phi(q)$ is a known geometric function and $\delta_{\pi \pi}(p)$ is the s-wave $\pi \pi \to \pi \pi$ scattering phase shift due only to QCD. This relation holds up to neglected, exponentially suppressed corrections of the form $e^{- M_\pi L}$. In order to extract a physically meaningful decay amplitude directly, one demands that the final two-pion state has the energy of the initial kaon. In the center-of-mass frame this requires tuning the box size, so that one of the energy levels coincides with the kaon mass, $E_k(L)=M_K$. 

This result has since been generalized to states with non-zero total momentum in the finite-volume frame, multiple coupled two-particle channels with non-identical and non-degenerate particles, to particles with intrinsic spin, and to transitions in which the current carries angular-momentum, momentum and energy, so that these quantum numbers differ between the initial and final states \cite{KSS2005,Christ2005,MeyerTimelike2011,HSmultiLL,BricenoTwoPart2012,BHWOneToTwo,BHOneToTwoSpin}.

In the present work we do not require the precise definition of $\phi(q)$. For our purposes it is sufficient to note that
\begin{equation}
\label{eq:infvolLL}
\lim_{L \to \infty} \mathcal C_k = \nu_k \,,
\end{equation}
where $\nu_k$ is the degeneracy of the $k$th state in the non-interacting theory, with the counting appropriate to non-identical particles, i.e.~$\nu_0=1$, $\nu_1=6, \cdots $. To see this, we first recall $4 \pi^2 q_k^2 / \phi'(q_k) = \nu_k$ where $q_k$ is the dimensionless momentum of the $k$th non-interacting state \cite{Lellouch2000}.%
\footnote{In the zero-momentum frame the non-interacting energies take the form $E_k(L) = 2 \sqrt{M_\pi^2 + \textbf n^2 (2 \pi/L)^2}$ corresponding to $q_k^2 = \textbf n^2$ where $\textbf n$ is a three-vector of integers.} % 
Equation (\ref{eq:infvolLL}) then follows from the fact that, as $L \to \infty$, any given finite-volume level will coincide with its non-interacting counterpart. In addition, the second, $\delta_{\pi \pi}$-dependent term is suppressed for fixed $k$ and large $L$ and thus vanishes in the infinite-volume limit.

Substituting Eq.~(\ref{eq:LL}) into the definition of the finite-volume spectral function, Eq.~(\ref{eq:FVSE}), we find
\begin{equation}
\rho_{\mathcal Q, \textbf 0}(E, \textbf 0, L) \equiv   \sum_{k}      \frac{\mathcal C_k}{2    E_k(L)^2 L^3} \   \big  \vert A_{K \to \pi \pi} [E_k(L)]  \big \vert^2 \ 2 \pi \delta \big( E - E_k(L) \big ) \,,
\end{equation}
and applying the smearing procedure then gives
\begin{equation}
\widehat \rho_{\mathcal Q, \textbf 0}(E, \textbf 0, L, \Delta) \equiv   \sum_{k}      \frac{\mathcal C_k}{2    E_k(L)^2 L^3} \   \big \vert A_{K \to \pi \pi} [E_k(L)]  \big \vert^2 \ 2 \pi \widehat \delta_{\Delta} \big( E , E_k(L) \big ) \,.
\end{equation}
Here we restrict our attention to the case where the incoming kaon and the the outgoing two-pion state both have zero spatial momentum.

The procedure outlined in Sec.~\ref{sec:FGR} dictates that one must now perform the $L \to \infty$ limit followed by $\Delta \to 0$ in order to recover the infinite-volume spectral function and from this the kaon width. We begin with the infinite-volume limit in isolation
\begin{align}
\lim_{L \to \infty} \widehat \rho_{\mathcal Q, \textbf 0}(E, \textbf 0, L, \Delta) & =  \lim_{L \to \infty}   \frac{1}{2}  \frac{1}{L^3} \sum_{\textbf k} \frac{1}{(2 \omega_{\textbf k})^2}    \     \big \vert A_{K \to \pi \pi} [2 \omega_{\textbf k}] \big  \vert^2 \ 2 \pi \widehat \delta_{\Delta} \big( E , 2 \omega_{\textbf k} \big ) \,, \\ &  = \frac{1}{2}  \int \frac{d^3 \textbf k}{(2 \pi)^3 (2 \omega_{\textbf k})^2}  \     \big \vert A_{K \to \pi \pi} [2 \omega_{\textbf k}] \big  \vert^2 \ 2 \pi \widehat \delta_{\Delta} \big( E , 2 \omega_{\textbf k} \big ) \,.
\end{align}
In the first step we used Eq.~(\ref{eq:infvolLL}) and then replaced $\sum_k \nu_k \to \sum_{\textbf k}$. We also used that, for a given state, the difference between $E_k(L)$ and the corresponding $2 \omega_{\textbf k}$ must vanish as $L \to \infty$. In the second step we replaced the sum over finite-volume momenta with an integral. This is justified for smooth integrands and in particular relies on the fact that the integrand depends only on smoothed delta functions with finite width $\Delta$.

At this stage we can send $\Delta \to 0$ and see that we recover the total width as is guaranteed by the general formalism presented in the previous section
\begin{align}
\Gamma_{K \to \pi \pi} & = \frac{1}{2 M_K} \lim_{\Delta \to 0}  \lim_{L \to \infty}    \widehat \rho_{\mathcal Q, \textbf 0}(M_K, \textbf 0, L, \Delta) \,, \\[5pt]
& =  \frac{1}{4 M_K}  \int \frac{d^3 \textbf k}{(2 \pi)^3 (2 \omega_{\textbf k})^2}  \     \vert A_{K \to \pi \pi}   \vert^2 \ 2 \pi  \delta \big( M_K -  2 \omega_{\textbf k} \big ) \,, \\[5pt]
& =  \frac{\sqrt{M_K^2/4 - M_\pi^2} }{16 \pi  M_K^2}       \vert A_{K \to \pi \pi}   \vert^2   \,,
\end{align}
where $A_{K \to \pi \pi}$ indicates the physical decay amplitude, in which the outgoing pion pair carries energy $M_K$.

We see that, in contrast to the standard Lellouch-L\"uscher approach in which one calculates the finite-volume matrix element and converts it to $A_{K \to \pi \pi}$ via Eq.~(\ref{eq:LL}), in this approach one removes the factor of $d \delta_{\pi \pi}(p)/ d p$ by sending $L \to \infty$.  This is only possible given the utility of the Backus-Gilbert method for directly extracting the smeared spectral function, together with the observation that the latter has a well-defined infinite-volume limit. In the case of a single two-particle decay channel, it is unlikely that this approach will be competitive with the Lellouch-L\"uscher method. However, as more channels open, our result continues to provide a viable method for estimating total decay widths. 

We have performed this same exercise with the generalization of the Lellouch-L\"uscher relation to coupled two-particle channels \cite{HSmultiLL,BHWOneToTwo,BHOneToTwoSpin}, and find that the expected result emerges in a similar manner. In the coupled-channel case the squared finite-volume matrix element can be expressed as a vector product of the infinite-volume matrix elements
\begin{equation}
\vert M_{k, D \to \mathcal Q} (L) \vert^2  =  \frac{1}{2  M_D L^9}   \  A_{D \to \alpha} (E_k)  \ \mathcal R_{\alpha \beta}(E_k, L)  \ A_{\beta \to D} (E_k ) \,,
\end{equation}
where $A_{D \to \alpha}(E) = \langle E,\, \alpha,\, \mathrm{out} \vert \mathcal H_{\mathcal Q}(0) \vert D  \rangle$ and $A_{\beta \to D}(E) = \langle D  \vert \mathcal H_{\mathcal Q}(0) \vert E,\,  \beta,\, \mathrm{in} \rangle$, with $\alpha$ and $\beta$ representing all quantum numbers of the two-particle asymptotic states. Deriving the corresponding smeared spectral function, and applying the infinite-volume limit, we find that the matrix becomes diagonal with each entry given by the degeneracy of the corresponding non-interacting state together with various kinematic factors. One thus perfectly recovers the integral over all three-momenta and the sum over channels, weighted with the proper factors. Sending $\Delta \to 0$ gives the total decay rate, equal to the individual rates summed over all open two-particle channels.

We close this section by noting that one might combine the Lellouch-L\"uscher relations with our approach in order to improve the smoothed spectral function. In particular, if the S-matrix is known in the two-particle sector, then one might fit the first few exponentials in $\widetilde G_{\mathcal Q, \textbf P}(\tau, \textbf p, L)$, subtract these states, and replace them with an integrated contribution of the infinite-volume matrix elements. Extracting $\widehat \rho_{\mathcal Q, \textbf P}(E, \textbf p, L)$ from this modified correlation function would give a flatter extrapolation towards $L \to \infty$ and $\Delta \to 0$, improving the precision of the extracted widths and differential rates.

\section{A numerical test case\label{sec:num}}

This section is devoted to illustrating, by means of a  numerical 
example, to what extent our procedure is able to reproduce infinite-volume total decay widths. The input data will consist of a Euclidean
correlation function $\widetilde G_{\mathcal Q, {\textbf 0}}(\tau_i,
{\bf 0}, L)$ of the type presented in Eq.~(\ref{eq:fourpoint}),
evaluated at $N_\tau$ discrete Euclidean time slices up to a maximum extent $L$, at which point we assume that the signal is lost. In this example we ignore finite-temperature effects, i.e.~we take the Euclidean temporal direction to have infinite extent.

We study a toy theory with three scalar particles
  denoted by, $\pi$, $K$ and $\phi$, with physical masses $M_\pi, M_K,
  M_\phi$, respectively, satisfying the hierarchy
\begin{equation}
3 M_\pi < 2 M_K < M_\phi  \,,
\end{equation}
and with interactions given by
\begin{equation}
\mathcal L(x) \supset  \frac{\lambda}{6} \phi(x) \pi(x)^3 + \frac{g M_\phi}{2} \phi(x) K(x)^2   \,.
\end{equation}
Treating these interactions perturbatively,
our goal is, given the Euclidean correlator, to reproduce the complete spectral function
$\rho_{\mathcal Q, \textbf 0}(\omega, \textbf 0)$, which, when evaluated
at $\omega = M_\phi$, gives access to the total decay width of the $\phi$ particle 
to leading order in the dimensionless couplings $\lambda$ and $g$.
We will neglect final-state interactions altogether.

\subsection{Finite Volume}
Before turning to the inverse problem of calculating $\widehat
\rho_{\mathcal Q, \textbf 0}(\overline \omega, \textbf 0, L, \Delta)$, we
need to construct the finite-volume Euclidean correlator that will
serve as input data. To proceed, note that the finite-volume matrix
element [Eq.~(\ref{eq:FVME2})] between a $\phi$-state and a two-$K$ state projected to $\bf p = \bf 0$ is given by
\begin{equation}
 \vert M_{k, \phi \to KK}(\textbf 0, L) \vert^2  = g^2 M_\phi^2  \frac{ \nu_k}{4   M_\phi E_k(L)^2 L^9}  \bigg \vert_{E_k(L) = 2 \sqrt{M_K^2 + (2 \pi/L)^2 \textbf q_k^2}} \,,
\end{equation}
where $\textbf q_k^2 = k$ starting with $k=0$, and where $\nu_k$ is the number of integer vectors $\textbf q$, satisfying $\textbf q^2 = k$.  For example $\nu_0 = 1$ and $\nu_1=6$. Following Eq.~(\ref{eq:GtildeCompset}) leads to a correlation function of the form
\begin{equation}
\frac{1}{M_\pi^3}\widetilde G_{KK}(\tau_i, {\bf 0}, L)  = \frac{g^2 M_\phi^2}{2 (M_\pi L)^3} \sum_k   \frac{ \nu_k}{    E_k(L)^2  } e^{- E_k(L) \tau_i } \bigg \vert_{E_k(L) = 2 \sqrt{M_K^2 + (2 \pi/L)^2 \textbf q_k^2}} \,,
\end{equation}
where we have divided by $M_\pi^3$ to give a dimensionless quantity. Analogously, one obtains for the three-$\pi$ amplitude
\begin{equation}
\vert M_{k, \phi \to \pi \pi \pi } (\textbf 0, L) \vert^2  = \lambda^2  \frac{ \nu_k}{12   M_\phi 2 \omega_{\textbf n_k} 2 \omega_{\textbf m_k} 2 \omega_{\textbf n_k + \textbf m_k} L^{12}}\,,
\end{equation}
where $\textbf n_k , \textbf m_k \in \mathds{Z}^3$ are integer vectors representing the $k$th state and
\begin{equation}
\omega_{\textbf n} =  \sqrt{M_\pi^2 + (2 \pi/L)^2 \textbf n^2 } \,.
\end{equation}
For example, the $k=0$ ground state is represented by $\textbf n=\textbf m = 0$ corresponding to $E_0(L) = 3M_\pi$ with degeneracy $1$. The next state can be represented with $\textbf n = - \textbf m$, with $\vert \textbf n \vert = \vert \textbf m \vert = 1 $, and has degeneracy $\nu_1=18$. The Euclidean correlator can then be written as
\begin{equation}
\frac{1}{M_\pi^3}\widetilde G_{\pi \pi \pi}(\tau_i, {\bf 0}, L) = \frac{\lambda^2}{48 M_\pi^3 L^6} \sum_k  \frac{\nu_k}{\omega_{\textbf n_k} \omega_{\textbf m_k} \omega_{\textbf n_k + \textbf m_k}} e^{- E_k(L) \tau_i } \bigg \vert_{E_k(L) =  \omega_{\textbf n_k} + \omega_{\textbf m_k} + \omega_{\textbf n_k + \textbf m_k}}\, .
\end{equation}
The full correlator that serves as input for the Backus-Gilbert procedure is then
\begin{equation}
\widetilde G_{\mathcal Q}(\tau_i, {\bf 0}, L) = \frac{1}{M_\pi^3}\widetilde G_{KK}(\tau_i, {\bf 0}, L) + \frac{1}{M_\pi^3}\widetilde G_{\pi \pi \pi }(\tau_i, {\bf 0}, L)\, .
\end{equation}

\subsection{Infinite Volume}
Since we want to compare the outcome of the Backus-Gilbert spectral-function reconstruction to the
exact infinite-volume result, we briefly derive the infinite-volume
contributions to the total decay width for both the two-$K$ and
three-$\pi$ channels. Starting with $\phi \to K K$, the decay
amplitude is given by
\begin{equation}
\mathcal A_{\phi \to K K} = g M_\phi \,.
\end{equation}
The main point here is that the amplitude is energy-independent -- the mass $M_\phi$ appears only
because we have chosen to parametrize the interaction by a dimensionless coupling constant $g$. 
Integration over the two-particle phase space yields a decay width
\begin{equation}
\frac{\Gamma_{\phi \to K K}}{M_\pi} =  \frac{g^2}{32 \pi} \frac{M_\phi}{M_\pi}   \sqrt{1 - \frac{4 M_K^2}{M_\phi^2}}      \,.
\end{equation}
Analogously, the three-$\pi$ decay amplitude is energy-independent,
\begin{equation}
\mathcal A_{\phi \to \pi \pi \pi} = \lambda \,.
\end{equation}
Integrating over the three-particle phase space yields 
\begin{equation}
\frac{\Gamma_{\phi \to \pi \pi \pi}}{M_\pi} = \frac{\lambda^2}{3072 \pi^3} \frac{M_\phi}{M_\pi} \mathcal F(M_\phi/M_\pi) \,.
\end{equation}
Here $\mathcal{F}(M_\phi/M_\pi)$, shown in the left pannel of Fig.~\ref{fig:resolution}, measures the reduction of phase space relative to the case of $M_\pi=0$ where the decay products are massless. The definition is
\begin{equation}
\mathcal F(x) \equiv \frac{2}{x^4}   \int_{4}^{(x-1)^2} d m_{12}^2 \ \int_{m_{23, - }^2}^{m_{23, +}^2} d m_{23}^2 \,,
\end{equation}
where
\begin{equation}
m_{23, \pm}^2 \equiv  \frac{x^2 - m_{12}^2 +3}{2} \pm \frac{1}{2}  \left [ \big (m_{12}^2 - 4 \big)  \left ( \frac{\left(x^2-1\right)^2}{m_{12}^2 }-2 (x^2+1 )+ m_{12}^2  \right ) \right ]^{1/2}   \,.
\end{equation}
Note in particular that $\mathcal F(3) = 0$ and $\mathcal F(\infty)=1$. This function has no simple analytic form, but can easily be calculated numerically to arbitrary precision. 

Finally, since we want to reconstruct the full spectral function, it is instructive to write the full infinite-volume result for every $\omega$,
\begin{equation}
\frac{1}{2 M_\phi M_\pi} \rho_{\mathcal Q, \textbf 0}(\omega, \textbf 0) =  \frac{\lambda^2}{3072 \pi^3} \frac{M_\pi}{M_\phi} \left(\frac{\omega}{M_\pi}\right)^2  \mathcal F(\omega/M_\pi) \theta(\omega-3M_\pi) + \frac{g^2}{32 \pi} \frac{M_\phi}{M_\pi}   \sqrt{1 - \frac{4 M_K^2}{\omega^2}} \theta(\omega-2M_K)\, .
\end{equation}

\subsection{Inverse problem and smoothing procedure: Backus Gilbert reconstruction}
The aim is to recover the full spectral function
\begin{equation}
\rho_{\mathcal Q, \textbf 0}(\overline{\omega}, \textbf 0) = \lim_{\Delta \to 0}  \lim_{L \to \infty}    \widehat \rho_{\mathcal Q, \textbf 0}(\overline \omega, \textbf 0, L, \Delta) \,,
\end{equation}
where in particular for $\overline{\omega} = M_\phi$ we get an estimate for the total decay width
\begin{equation}
\frac{1}{2 M_\phi M_\pi} \lim_{\Delta \to 0}  \lim_{L \to \infty} \widehat \rho_{\mathcal Q, \textbf 0}(M_\phi, \textbf 0, L, \Delta) = \frac{\Gamma_{\phi \to K K}}{M_\pi} + \frac{\Gamma_{\phi \to \pi \pi \pi}}{M_\pi} \,.
\end{equation}
For the numerical application, we choose $M_K/M_\pi=3.55$ and $M_\phi/M_\pi=7.30$,
$g=1$ and $\lambda  = 10\sqrt{8}$. We use as input points $\tau_i=i\cdot a$, with $aM_\pi = 0.066$ and $1 \leq i \leq N_\tau$.

\begin{figure}
\begin{center}
\includegraphics[height=0.35\textwidth]{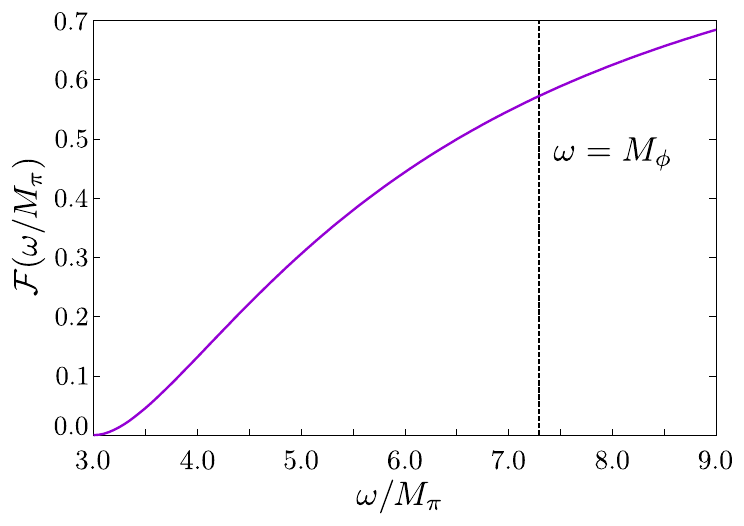} \includegraphics[height=0.35\textwidth]{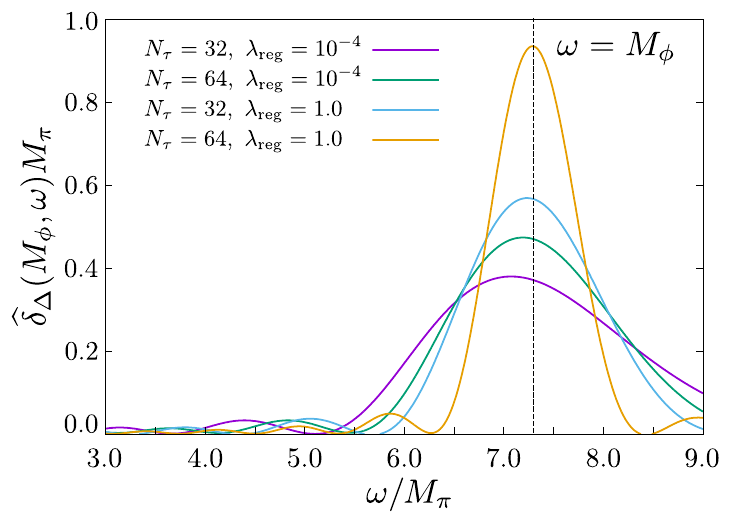}
\caption{Left: The function $\mathcal F(\omega/M_\pi)$ gives the three-body phase space with respect to the massless case. The function interpolates from 0 at three-particle threshold [$\mathcal F(3)=0$] to unity at infinite $\omega$ [$\lim_{\omega \to \infty} \mathcal F(\omega/M_\pi) = 1$]. Right: Various resolution functions, plotted as a function of $\omega$ with $\overline \omega= M_\phi$. The resolution functions with $\lambda_\text{reg}=10^{-4}$ were used in the analysis. The unregulated curves, with $\lambda_\text{reg}=1.0$, show how the Backus-Gilbert method converges to sharply peaked resolution functions as $N_\tau$ is increased. \label{fig:resolution}}
\end{center}
\end{figure}

Following the discussion of Sec.~\ref{sec:FGR}, we construct a family of resolution functions $\widehat{\delta}_\Delta(\overline{\omega},\omega)$ by finding the optimal coefficients $C_i(\overline{\omega},\Delta)$, $(i=1,\dots,N_\tau)$, that minimize the width subject to the area constraint of Eq.~(\ref{eq:area}). The width is controlled by the number of points, as can be seen from Fig.~\ref{fig:resolution}, where we give examples of resolution functions used in the analysis, centered at ${\omega}=M_\phi$. The corresponding estimator $\widehat \rho_{\mathcal Q, \textbf 0}(\overline{\omega}, \textbf 0)$ is calculated via Eq.~(\ref{eq:rhohatest}). Since Backus-Gilbert is a linear method, an error estimate on $\widehat \rho_{\mathcal Q, \textbf 0}(\overline{\omega}, \textbf 0)$ has to come through the covariance matrix of the input data $\widetilde G_{\mathcal Q}(\tau_i, {\bf 0}, L)$. We take a realistic covariance matrix, $S_{ij}$, from a pseudoscalar meson two-point function calculated on an $N_f=2$ CLS ensemble, that we have used in the past in a similar context. (See Appendix C of Ref.~\cite{Brandt:2015sxa}.) We scale the covariance matrix to give the same relative uncertainty (about $2\%$) on the toy correlator, $\widetilde G_{\mathcal Q}(\tau_i, {\bf 0}, L)$, as was observed on the actual lattice data. For $N_\tau = 64$ we take the entire matrix and for $N_\tau=32$ we take the block corresponding to source-sink separations, $\tau$, satisfying $32 a \leq \tau < 64 a$.

In order to determine the optimal coefficients, $C_i(\overline{\omega},\Delta)$, a poorly conditioned matrix,
\begin{equation}
W_{ij}(\overline{\omega}) = \int^\infty_0 d\omega e^{-\omega \tau_i} (\omega-\overline{\omega})^2 e^{-\omega \tau_j} \,,
\end{equation} 
must be inverted (ideally with high precision). The error on $\widehat \rho_{\mathcal Q, \textbf 0}(\overline{\omega}, \textbf 0)$ is kept under control by making the replacement $W_{ij}(\overline{\omega}) \to \lambda_\text{reg} W_{ij}(\overline{\omega}) + (1-\lambda_\text{reg})S_{ij}$. In this way the magnitudes of the coefficients $C_{i}(\overline{\omega},\Delta)$, which otherwise exhibit large oscillations as a function of $i$, are tamed, and the statistical error on $\widehat \rho_{\mathcal Q, \textbf 0}(\overline{\omega}, \textbf 0)$ can be kept under control. In the present example, we aim for a precision of $5-10\%$ and find that this is achieved with $\lambda_\text{reg}\sim 10^{-4}$. The regulation parameter, $\lambda_\text{reg}$, parametrizes a trade-off between resolving power (smaller $\Delta$) and statistical error. For a more detailed explanation on the choice of $\lambda_\text{reg}$ see the discussion in Sec.~IV E of Ref.~\cite{Brandt:2015sxa}.

\begin{figure}[h!]
\begin{center}
\includegraphics[height=0.35\textwidth]{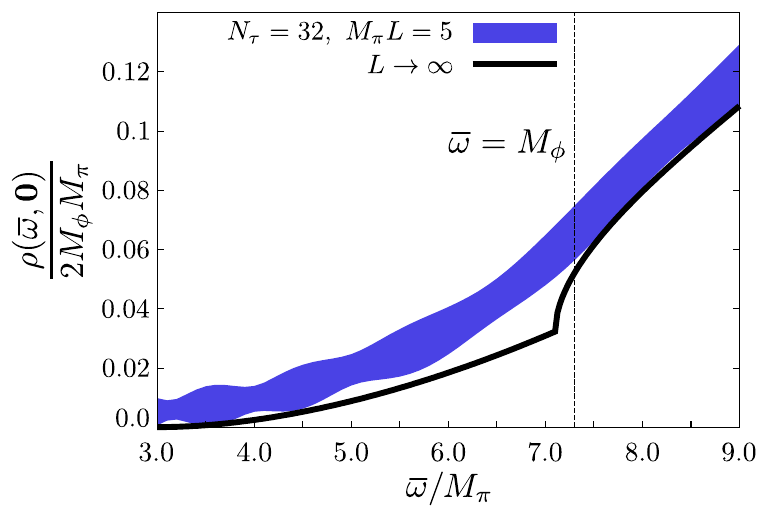}
\includegraphics[height=0.35\textwidth]{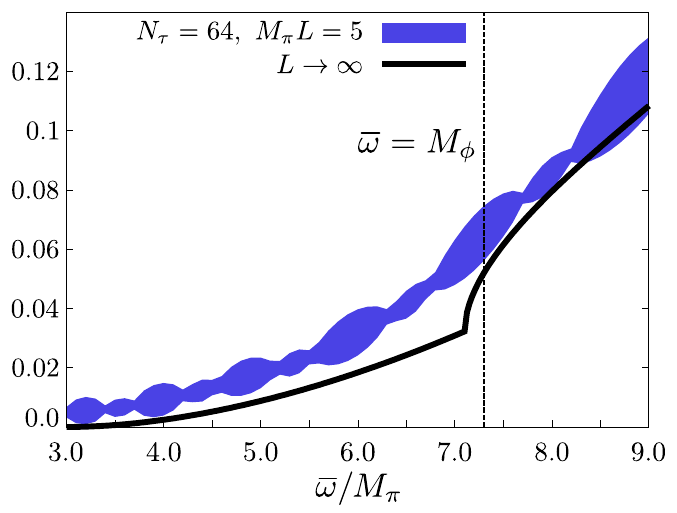}
\includegraphics[height=0.35\textwidth]{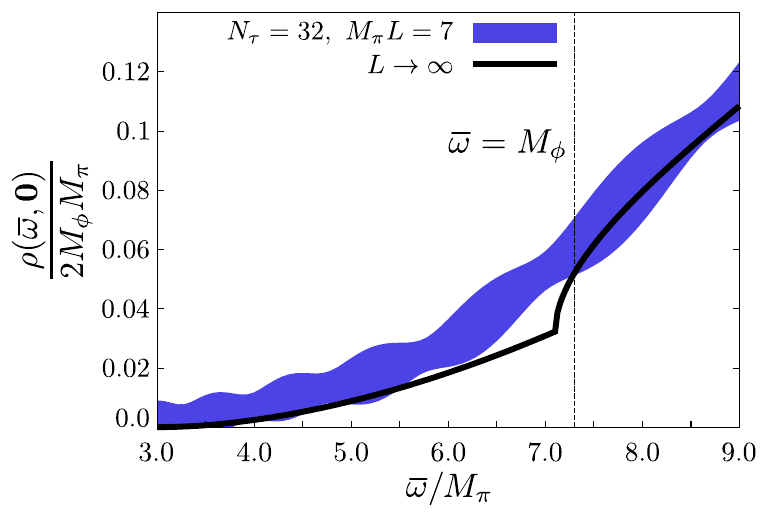}
\includegraphics[height=0.35\textwidth]{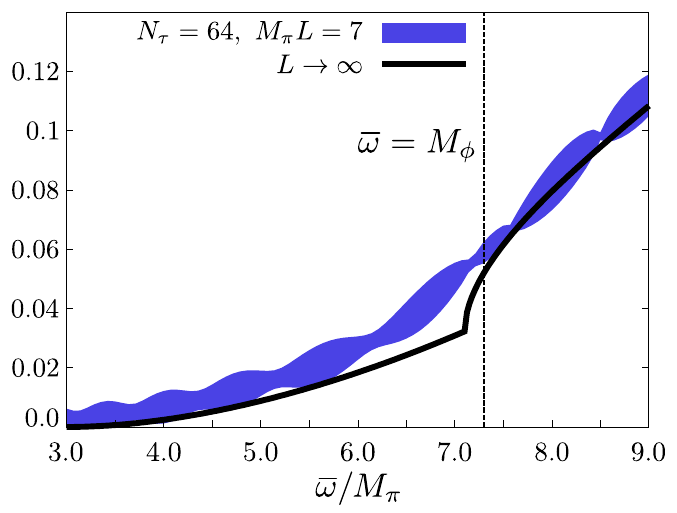}
\includegraphics[height=0.35\textwidth]{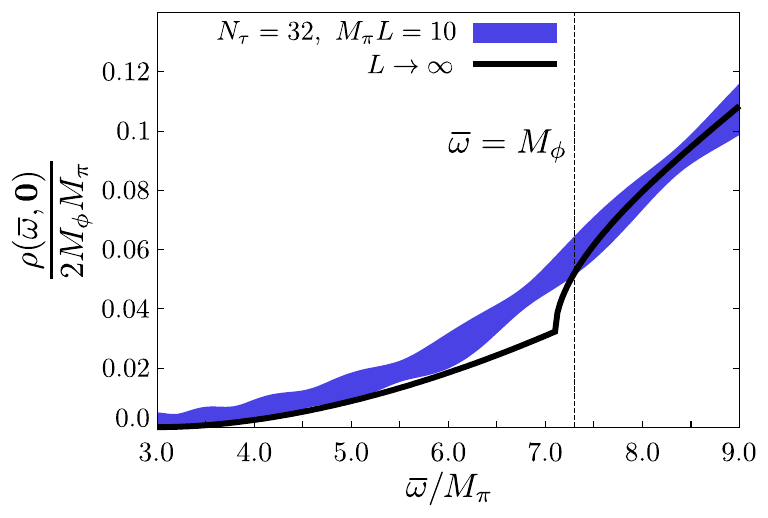}
\includegraphics[height=0.35\textwidth]{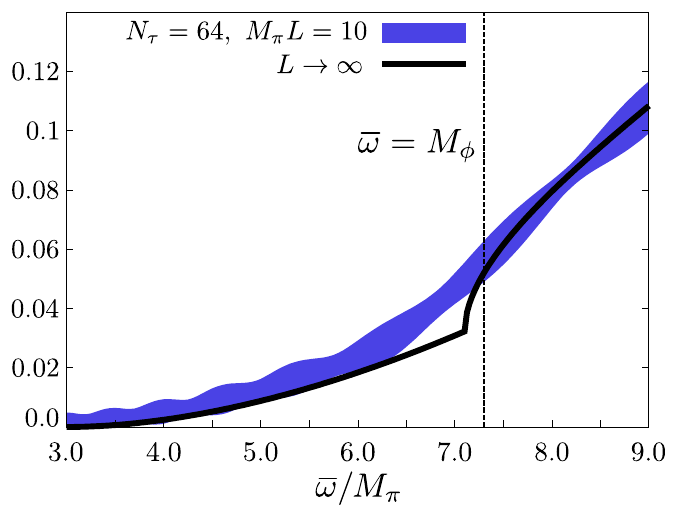}
\caption{Output of the regulated Backus-Gilbert algorithm for different values of $M_\pi L = 5,7,10$ and $N_\tau = 32, 64$. The ratios of $M_\phi/M_\pi, M_K/M_\pi$ and the coupling constants are given in the text. \label{fig:rhoresults}}
\end{center}
\end{figure}

The results of our Backus-Gilbert analysis are summarized in Fig.~\ref{fig:rhoresults}. As can be seen from the figure, the agreement of the Backus-Gilbert result with the infinite-volume spectral function is reasonable and improves for increasing $M_\pi L$ as expected. If one instead tries to reduce $\Delta$ at constant $M_\pi L$, the result becomes unstable since the finite-volume energy levels start to be resolved individually and the result is eventually dominated by finite-volume effects. This illustrates the importance of the order of the limits $L \to \infty$ and $\Delta \to 0$. Nevertheless, the window of reasonable $\Delta$ for a given $L$ seems to be large enough that it may well be useful in realistic numerical lattice applications. As previously mentioned, the resolution function from the Backus-Gilbert method %and used in the analysis of lattice data 
can be used to smoothen the experimental data in the case of differential decay rates. This procedure removes the uncertainty inherent in the solution of the inverse problem at the cost of comparing a somewhat less differential quantity.

\section{Conclusions\label{sec:conc}}

In this work we have introduced a new method for directly determining hadronic decay widths and transition rates for semi-leptonic scattering and decay processes. The central advantage of our approach is that it can accommodate final states with any number of hadrons. 

As we detail in Sec.~\ref{sec:FGR}, our idea is to construct a Euclidean correlator such that the corresponding spectral function directly gives the decay width or transition rate of interest. We then propose applying the Backus-Gilbert method, which gives an estimator of the finite-volume spectral function, smeared by a known resolution function $\widehat \delta_{\Delta}(\overline \omega, \omega)$ of width $\Delta$. Taking the limit $L \to \infty$ followed by $\Delta \to 0$ then directly gives the experimental observable. As we discuss in Sec.~\ref{sec:FGR}, and illustrate with a toy example in Sec.~\ref{sec:num}, in a realistic numerical calculation we expect that reducing $1/L$ and $\Delta$ along a suitable trajectory will provide a good estimate of the target infinite-volume unsmeared spectral function.

Although we have focused on the Backus-Gilbert method, we would like to stress that a vast number of different methods exist for solving the inverse Laplace problem. Any of these may be applied in the formalism presented in this work, provided that one can define a width, $\Delta$, that can be varied together with the box size, $L$, to estimate the double limit. Indeed, even within the Backus-Gilbert approach one has freedom to adjust the regulation parameter, $\lambda_\text{reg}$, introduced in Sec.~\ref{sec:num} as well as the minimization condition given in Eq.~(\ref{eq:mincon}). We have applied the Backus-Gilbert method on vacuum correlators in Refs.~\cite{Brandt:2015aqk,Brandt:2015sxa}.

The central advantage of the Backus-Gilbert approach is that it offers a model-independent, unbiased estimator of the smeared, finite-volume spectral function with a precisely known resolution function that is independent of the correlator data. For every value of the final-state energy that one aims to extract, one may vary all inputs into the Backus-Gilbert approach to design the optimal resolution function given the quality of the data available. Nonetheless, it may often be the case that the difficulty of reducing the width $\Delta$ is the limiting factor of the calculation. To this end we re-emphasize that one may also smear experimental or model data with the same resolution function to perform a fully controlled comparison.

In certain cases the resolution function may wash away features that one aims to extract. While this is undesirable, the method does faithfully return the information contained in the Euclidean correlator data about the width or rate of interest. Indeed, in the context of constraining scattering amplitudes via
L\"uscher's quantization condition, it is in practice well known that a
single correlation function provides limited information on
finite-volume energies. The modern approach here is to instead construct a matrix of correlators, with a large operator basis, and diagonalize this in order to reliably extract the excited finite-volume states. Analogous methods in the context of finite-volume spectral functions are under investigation \cite{Harris:2016usb}.

\acknowledgements{We thank all our colleagues in the Mainz lattice group for helpful discussions, encouragement, and support. We additionally acknowledge Steve Sharpe, Keh-Fei Liu and Shoji Hashimoto for helpful comments on the first version of this manuscript.  D.~R.~wishes to specially thank Guy Moore for very interesting discussions, and  M.~T.~H.~would like to thank Steve Sharpe and Ra\'ul Brice\~no for insightful discussions and helpful feedback. This work was supported in part by DFG grant ME 3622/2-2 and by the State of Hesse.}

% commented out so v.4-1 style is used
%\bibliographystyle{apsrev} %%% physical review
\bibliography{FGR} %%% ref.bib file

\end{document}